%% file: main.tex
\documentclass[amsmath,amssymb,twocolumn,notitlepage,nofootinbib,superscriptaddress,floatfix,eprint,aps]{revtex4-2}

\usepackage[utf8]{inputenc}
\usepackage[english]{babel}
\usepackage{graphicx}
\usepackage{float}
\usepackage{color}
\usepackage{tabularx}
\usepackage[usenames,dvipsnames,table]{xcolor}
\usepackage[normalem]{ulem}
\usepackage{amsthm}
\usepackage{stmaryrd}
\usepackage{mathtools}
\usepackage{eucal}
\usepackage{times,bbm,amsmath,amssymb}
\usepackage{epsfig,color}
\usepackage{xcolor}
\usepackage{bm}
\usepackage{booktabs}
\usepackage{float,siunitx}
\usepackage{booktabs}

\definecolor{quantumviolet}{HTML}{53257F}
\definecolor{navy}{RGB}{47,60,126}
\definecolor{darkviolet}{RGB}{99,56,142}
\definecolor{darkgreen}{RGB}{39,174,96}

\usepackage{sidecap}
\usepackage[scaled=.8]{couriers}
\usepackage{pstricks}
\usepackage{multirow}
\usepackage{placeins}
\usepackage{relsize}
\usepackage{pst-grad,bm}
\usepackage{epigraph}
\usepackage{gensymb}
\usepackage{longtable}
\usepackage{soul}
\usepackage{ulem}
\usepackage{xr-hyper}
\usepackage{hyperref}
\usepackage[nameinlink]{cleveref} 
\crefname{section}{Section}{Sections}
\crefname{equation}{Eq.}{Equations}
\crefname{figure}{Fig.}{Figures}
\crefname{table}{Table}{Tables}
\crefname{appendix}{Appendix}{Appendices}
\crefname{theorem}{Theorem}{Theorems}
\crefname{thm}{Theorem}{Theorems}
\crefname{cor}{Corollary}{Corollaries}
\crefname{lemma}{Lemma}{Lemmas}
\crefname{proposition}{Proposition}{Propositions}
\crefname{definition}{Definition}{Definitions}
\crefname{algorithm}{Algorithm}{Algorithms}
\hypersetup{
    colorlinks=true,
    linkcolor=navy,
    citecolor=quantumviolet,
    urlcolor=quantumviolet,
    breaklinks=true,
}

\let\autoref\cref

\usepackage{layouts}
\usepackage{acronym}
\usepackage{physics}
\usepackage{easyReview}

\usepackage{soul}

\usepackage{amsmath,amssymb,amsfonts}
\usepackage{textcomp}
\usepackage{physics}

\usepackage[ruled,linesnumbered]{algorithm2e}
\usepackage{verbatim}
\usepackage{tabularx}

\usepackage{booktabs}

\usepackage{microtype}
\microtypecontext{spacing=nonfrench}
\microtypesetup{
protrusion={true,nocompatibility},
activate={true,nocompatibility},
tracking=true,
kerning=true,
spacing={true}
}
\usepackage[all=normal,floats=tight,mathspacing=tight,wordspacing=tight,paragraphs=normal,tracking=tight,charwidths=tight,mathdisplays=tight,sections=normal,margins=normal]{savetrees}

\usepackage[style=american,autopunct=true]{csquotes}

\usepackage{placeins}

\newtheorem{definition}{Definition}

\begin{document}

\title{Photonic Quantum Convolutional Neural Networks with Adaptive State Injection}

\author{ Léo Monbroussou}
\affiliation{Laboratoire d’Informatique de Paris 6, CNRS, Sorbonne Université, 75005 Paris, France}
\affiliation{CEMIS, Direction Technique, Naval Group, 83190 Ollioules, France}

\author{Beatrice Polacchi}
\affiliation{Dipartimento di Fisica, Sapienza Universit\`{a} di Roma, P.le Aldo Moro 5, I-00185 Roma, Italy}

\author{Verena Yacoub}
\affiliation{Laboratoire d’Informatique de Paris 6, CNRS, Sorbonne Université, 75005 Paris, France}

\author{Eugenio Caruccio}
\affiliation{Dipartimento di Fisica, Sapienza Universit\`{a} di Roma, P.le Aldo Moro 5, I-00185 Roma, Italy}

\author{Giovanni Rodari}
\affiliation{Dipartimento di Fisica, Sapienza Universit\`{a} di Roma, P.le Aldo Moro 5, I-00185 Roma, Italy}

\author{Francesco Hoch}
\affiliation{Dipartimento di Fisica, Sapienza Universit\`{a} di Roma, P.le Aldo Moro 5, I-00185 Roma, Italy}

\author{Gonzalo Carvacho}
\affiliation{Dipartimento di Fisica, Sapienza Universit\`{a} di Roma, P.le Aldo Moro 5, I-00185 Roma, Italy}

\author{Nicol\`o Spagnolo}
\affiliation{Dipartimento di Fisica, Sapienza Universit\`{a} di Roma, P.le Aldo Moro 5, I-00185 Roma, Italy}

\author{Taira Giordani}
\affiliation{Dipartimento di Fisica, Sapienza Universit\`{a} di Roma, P.le Aldo Moro 5, I-00185 Roma, Italy}

\author{Mattia Bossi}
\affiliation{Dipartimento di Fisica, Politecnico di Milano, Piazza Leonardo da Vinci 32, 20133 Milano, Italy}
\affiliation{Istituto di Fotonica e Nanotecnologie, Consiglio Nazionale delle Ricerche (IFN-CNR), 
Piazza Leonardo da Vinci, 32, 20133 Milano, Italy}

\author{Abhiram Rajan}
\affiliation{Dipartimento di Fisica, Politecnico di Milano, Piazza Leonardo da Vinci 32, 20133 Milano, Italy}
\affiliation{Istituto di Fotonica e Nanotecnologie, Consiglio Nazionale delle Ricerche (IFN-CNR), 
Piazza Leonardo da Vinci, 32, 20133 Milano, Italy}

\author{Niki Di Giano}
\affiliation{Dipartimento di Fisica, Politecnico di Milano, Piazza Leonardo da Vinci 32, 20133 Milano, Italy}
\affiliation{Istituto di Fotonica e Nanotecnologie, Consiglio Nazionale delle Ricerche (IFN-CNR), 
Piazza Leonardo da Vinci, 32, 20133 Milano, Italy}

\author{Riccardo Albiero}
\affiliation{Istituto di Fotonica e Nanotecnologie, Consiglio Nazionale delle Ricerche (IFN-CNR), 
Piazza Leonardo da Vinci, 32, 20133 Milano, Italy}

\author{Francesco Ceccarelli}
\affiliation{Istituto di Fotonica e Nanotecnologie, Consiglio Nazionale delle Ricerche (IFN-CNR), 
Piazza Leonardo da Vinci, 32, 20133 Milano, Italy}

\author{Roberto Osellame}
\email{roberto.osellame@cnr.it}
\affiliation{Istituto di Fotonica e Nanotecnologie, Consiglio Nazionale delle Ricerche (IFN-CNR), Piazza Leonardo da Vinci, 32, 20133 Milano, Italy}

\author{Elham Kashefi}
\email{elham.kashefi@lip6.fr}
\affiliation{Laboratoire d’Informatique de Paris 6, CNRS, Sorbonne Université, 75005 Paris, France}
\affiliation{School of Informatics, University of Edinburgh, 10 Crichton Street, EH8 9AB Edinburgh, United Kingdom}

\author{Fabio Sciarrino}
\email{fabio.sciarrino@uniroma1.it}
\affiliation{Dipartimento di Fisica, Sapienza Universit\`{a} di Roma, P.le Aldo Moro 5, I-00185 Roma, Italy}

\begin{abstract}

Linear optical architectures have been extensively investigated for quantum computing and quantum machine learning applications. 
Recently, proposals for photonic quantum machine learning have combined linear optics with resource adaptivity, such as adaptive circuit reconfiguration, which promises to enhance expressivity and improve algorithm performances and scalability.
Moreover, linear optical platforms preserve some subspaces due to the fixed number of particles during the computation, a property recently exploited to design a novel quantum convolutional neural networks. This last architecture has shown an advantage in terms of running time complexity and of the number of parameters needed with respect to other quantum neural network proposals. 
In this work, we design and experimentally implement the first photonic quantum convolutional neural network (PQCNN) architecture based on {particle-number} preserving circuits equipped with state injection, an approach recently proposed to {increase the controllability of} linear optical circuits. Subsequently, we experimentally validate the PQCNN for a binary image classification on a photonic platform utilizing a semiconductor quantum dot-based single-photon source and programmable integrated photonic interferometers comprising 8 and 12 modes.
In order to investigate the scalability of the PQCNN design, we have performed numerical simulations on datasets of different sizes.
We highlight the potential utility of a simple adaptive technique for a nonlinear Boson Sampling task, compatible with near-term quantum devices.

\end{abstract}

\maketitle

\section{Introduction}
In 2001, E. Knill, R. Laflamme, and G. J. Milburn demonstrated that universal quantum computing can be achieved by using only beam splitters, phase shifters, single-photon sources, and photo-detectors \cite{knill2001scheme}. 
However, the resources needed to fully achieve such a task are demanding, and only noisy intermediate-scale quantum processors can be realized with current photonic technologies. As a consequence, this triggered the study of sub-universal platforms such as Boson Sampling \cite{aaronson2011computational}, which aims at showing a quantum computational advantage with the current technology, as demonstrated by the recent realizations of Gaussian Boson Sampling \cite{zhong2020quantum, Zhong2021, Madsen2022}. However, so far, it has been difficult to map useful problems in the Boson Sampling scheme. Nonetheless, the development of photonic quantum technologies \cite{flamini2018photonic,Psi_Quantum} and, in particular, integrated ones \cite{wang2020integrated,pelucchi2022potential, Giordani2023_rev}, encourages the search for algorithms and architectures that allow for {a useful} intermediate quantum computational advantage while being compatible with the available hardware. Indeed, photonic integrated interferometers allow for a large number of modes in a compact layout and for intrinsic phase stability, thus significantly increasing the capability of implementing large systems with respect to bulk interferometers.

A promising route to extend the expressivity of photonic architectures and the range of applications in the field of quantum machine learning \cite{biamonte2017quantum,schuld2014quest,ciliberto2018quantum,romero2017quantum,dunjko2018machine,biamonte2021universal}  is represented by the introduction of nonlinearity \cite{, spagnolo2023non, Steinbrecher2019} or particular adaptive operations \cite{chabaud2021quantum, hoch2025quantum}. This is exemplified by the theoretical work \cite{chabaud2021quantum} that proposes to apply Boson Sampling equipped with adaptive circuit reconfiguration, to tackle quantum machine learning tasks, for instance kernel estimation \cite{hoch2025quantum}.  
The so-called Adaptive Boson Sampling scheme \cite{chabaud2021quantum}, recently demonstrated in \cite{hoch2025quantum}, proposes to apply unitary transformations according to the measurement outcomes over subsets of photons and modes, which represent the adaptive parts of the protocol. Within this framework, an alternative approach was recently introduced in \cite{monbroussou2024quantum}, which substitutes adaptive evolutions with adaptive injection into the optical circuits of new photonic states. 

More precisely, state injection is a measurement-based operation that triggers the preparation and injection of new photons according to the outcomes of single-photon detections in selected channels. The resulting architecture alternates linear optical layers with adaptive state injections that can act as the nonlinear gadget of the protocol. Such an adaptive linear optic scheme enables a more straightforward and scalable mapping of optical neural networks \cite{Steinbrecher2019} to the operations of an adaptive photonic circuit. 

Various models for quantum learning attempted to expand the class of machine learning algorithms that can be encoded in quantum hardware \cite{schuld2014quest, dunjko2018machine, Cong2019}, for example, by looking at Classical Convolutional Neural Networks (CCNNs), currently widely used for image, audio, or video recognition \cite{LeCun_LeNet,Li2022, Alzubaidi2021}.
A quantum counterpart of CCNNs has been recently proposed in \cite{monbroussou2024subspace}, offering notable advantages in terms of scalability, trainability, and expressivity. 
In particular, this scheme introduces a Quantum Convolutional Neural Network (QCNN) architecture based on Hamming weight-preserving quantum circuits. The proposed architecture offers polynomial speed-ups over classical deep learning architectures and uses convolutional and measurement-based pooling layers to preserve quantum state symmetries while introducing nonlinearity.

In this work, we introduce and experimentally demonstrate an architecture for Photonic Quantum Convolutional Neural Networks (PQCNNs) which explores the state injection architecture presented in \cite{monbroussou2024quantum} to realize a photonic implementation of  {subspace} preserving QCNNs \cite{monbroussou2024subspace}. We realize a proof-of-concept experiment by employing a cutting-edge single-photon source based on a semiconductor Quantum Dot (QD) \cite{somaschi2016near,loredo2019generation}, a time-to-spatial demultiplexer, and universal programmable 12-mode and 8-mode interferometers realized with the femtosecond laser-writing technique \cite{Ceccarelli2020,Corrielli2021}.
The designed PQCNN scheme is tailored to the experimental platform at hand, with the goal of carrying out a binary image classification.
As a complement to the experimental investigation, we provide a systematic study on the scaling and complexity of the protocol, by leveraging numerical simulations on larger quantum systems, demonstrating the potential behind the proposed scheme for PQCNNs. Finally, we offer a tailor-made library that allows for efficient simulation of such architecture by performing the computation in the most appropriate subspaces.

\section{Photonic Quantum Convolutional Neural Networks}
\begin{figure}[h!]
    \centering
    \includegraphics[width=\columnwidth]{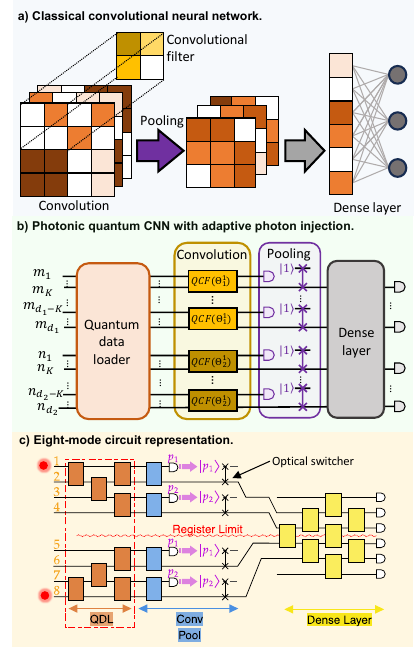}
    \caption{\textbf{Classical Convolutional Neural Networks (CCNN) and Photonic Quantum Convolutional Neural Networks (PQCNN) based on adaptive state injection. a)} CCNNs alternate convolutional and pooling layers to process the data features and reduce their dimensions. The last layer is a dense layer that connects the remaining features to the output nodes.
    \textbf{b)} PQCNN circuit representation for an input image of size $d_1 \times d_2$. The entire flow is composed of a Quantum Data Loader (QDL), a convolutional layer with a {Quantum Convolutional Filter (QCF)} of size $K \times K$, a pooling layer, and a dense layer before the detection. In the described scenario, the QDL takes as input two separated registers, indicated with $m_i$ and $n_i$, that encode, respectively, the rows and the columns of the input image. In the Pooling Layer, some modes are measured and, upon photon detection, a Fock state with one photon is injected into the adjacent mode. \textbf{c)} In this illustrative PQCNN involving eight modes, the QDL stage is performed by means of four Beam Splitters (BS) acting on the row and column registers respectively, while the convolutional layer is applied through a layer of four BSs on the input image. The distribution at the output of the pooling layer defines the probability of injecting extra photons into the dense layer. The latter is performed through eight BSs over six modes and a final readout layer.} 
    \label{fig:concept}
\end{figure}

The Classical Convolutional Neural Network (CCNN) is one of the most exploited deep learning architectures \cite{Li2022, Alzubaidi2021}, and has been part of significant achievements in many areas, including computer vision, time series analysis, and natural language processing. In its earliest demonstration \cite{LeCun_LeNet}, this architecture is typically composed of convolutional layers that extract important features from the data, pooling layers that reduce the size of the data during the computation, and a final dense neural network that mixes the features and performs the classification task.
A sketch of the CCNN architecture is shown in \autoref{fig:concept}a.
While CCNNs are highly effective in capturing spatial or local patterns in data, their performance can be further improved by introducing mechanisms that provide contextual awareness. 

In Ref.~\cite{monbroussou2024subspace}, a quantum counterpart of a CCNN is proposed, based on Hamming weight-preserving quantum circuits \cite{monbroussou2024trainabilityexpressivityhammingweightpreserving}. Hamming weight preservation shares strong similarities with photonic quantum circuits and, owing to its subspace-preserving properties \cite{diaz2023showcasing, Ragone2024, Fontana2023,Larocca2021, monbroussou2024trainabilityexpressivityhammingweightpreserving}, such an architecture is believed to mitigate the occurrence of Barren plateaus \cite{mcclean2018barren} -- a vanishing gradient phenomenon that commonly hampers the training of variational quantum circuits \cite{PRXQuantum.3.010313}.
Such a scheme aims to construct quantum stages similar to the ones of a typical CCNN. As shown in \autoref{fig:concept}b, the architecture of a Photonic Quantum Convolutional Neural Network (PQCNN) comprises of an initial quantum data loading layer that maps the classical image into a quantum register, while the remaining steps reproduce the prototypical CCNN operations, i.e. a sequence of convolutional, pooling, and dense layers.

Within the photonic approach proposed here, quantum data loading, convolutional, and dense layers are obtained through linear optics while also exploiting an adaptive state injection scheme in the pooling layer in order to enable dynamic adaptability and nonlinearity, as depicted in \autoref{fig:concept}b.
In the scheme, the pooling layer consists of performing single-photon detection in some modes, and, if a photon is detected, a Fock state with exactly one photon is injected in the next mode.
A particular instance of a PQCNN architecture involving an eight-mode circuit is illustrated in \autoref{fig:concept}c.
In the following subsections, we provide a detailed description of each stage of the proposed PQCNN architecture and of the specific instance that has been implemented in the experiment.

\subsection{Quantum Data Loader}\label{subsec:Tensor_Encoding}

The first procedure is Quantum Data Loading (QDL), an operative way to encode different data types in quantum states.
This algorithm is based on \textit{tensor encoding}, a particular case of amplitude encoding. Such an encoding maps the data features into the amplitude associated with reference-state vectors. In detail, let us consider a classical tensor of dimension $k$ such that $x = (x_{1, \dots, 1}, \dots,  x_{d_1, \dots, d_k}) \in \mathbb{R}^{d_1 \times \dots \times d_k}$. The corresponding photonic tensor-encoded state is:
    \begin{equation}\label{eq:Tensor_Encoding_modes}
        \ket{x} = \frac{1}{||x||} \sum_{i_1 \in [d_1]} \dots \sum_{i_k \in [d_k]}  x_{i_1, \dots, i_k} \ket{e_{d_1, i_1}} \otimes \dots \otimes \ket{e_{d_k, i_k}}
    \end{equation}
where $\ket{e_{d_l, i_l}} = \ket{0 \dots 0 1 0 \dots 0}$ represents a Fock state over $d_l$ modes, with a single excitation (photon) in mode $i_l$ and vacuum in all other modes. Therefore, the set $\left\{\ket{e_{d_l, i}} \mid i \in [d_l] \right\}$ represents a fixed family of $d_l$ orthonormal quantum states, while $||\cdot||$ denotes the $2$-norm of $\mathbb{R}^d$. Notice that the input state of the algorithm generally requires $m = \sum_{i \in [k]} d_i$ modes dispatched in $k$ different registers with a single particle in each of them. In \autoref{fig:concept}b, the QDL takes two-dimensional images in input. In this case, the first register, with modes $m_{d_1}$, represents rows, and the second, with modes $n_{d_2}$, represents the image columns.

To encode any tensor $x \in \mathbb{R}^{d_1 \times \dots \times d_k}$ within this framework, one needs to use a photonic architecture with $ m=\sum_{i=1}^k d_i$ modes that can freely control the amplitudes of the Fock states used for the tensor encoding in \autoref{eq:Tensor_Encoding_modes}. We note that considering $k$ photons distributed over $k$ registers, each constructed with an indipendent linear optical circuit spanning $d_j$ modes, constraints the tensor-encoded state of Eqn. (1) to be a separable state. In order to obtain a generic k-dimensional tensor in the proposed encoding, additional resources in terms of ancilla photons, modes or measurement feedforward would be required \cite{chabaud2021quantum,monbroussou2024quantum}.

\subsection{Convolutional Layer}\label{subsec:Conv_Layer}

The encoded data is then fed into a convolutional operation.
For $k$-dimensional tensor-encoded inputs, one needs again to consider $k$ separate registers of modes with only a single photon in each of them.
In the quantum circuits proposed in \cite{monbroussou2024subspace}, convolutional layers use the Reconfigurable Beam Splitter (RBS) gate {studied} in \cite{monbroussou2024trainabilityexpressivityhammingweightpreserving}, which applies a planar rotation between the states $\ket{01}$ and $\ket{10}$.
When using single photons, this operation can be directly performed with Beam Splitters (BS), another key feature making photonic platforms as the natural candidate for the proposed PQCNN architecture.
In \autoref{chap:RBS_Photonic_BS}, we detail the connection and the differences between BSs and RBS gates.
The convolutional layer described in \cite{monbroussou2024subspace} can be adapted to photonic platforms as follows: for each register and a {Quantum Convolutional Filter} (QCF) of size $K_1 \times \dots \times K_k$, the convolutional layer consists of applying the same circuit made of BSs to each partition of $K_i$ modes ($i \in \llbracket 1, k \rrbracket$) with the same set of variational parameters. This circuit {is represented in \autoref{fig:concept}b} for a $2$-dimensional tensor input, and a QCF of size $K \times K$. Notice that, as for classical convolutional layers, one can choose to use a convolutional layer of lower dimension with respect to the input dimension by keeping some registers unaffected on the circuit. The depth of the convolutional layer is $\mathcal{O}(K)$ with $K = \max(\{ K_1, \cdots K_k \})$ because only $K(K-1)/2$ parameters are needed to maximize the control of this circuit.
In \autoref{chap:Appendix_Convolutional_Layer}, we recall the definition of a classical convolutional layer and show that this photonic proposal performs an analogous operation. Namely, the output state is still tensor encoded, and thus the value of each new coefficient is a linear combination of the former values.

Consider for example a $d_1 \times d_2$ input tensor and a convolutional layer { with a $K \times K$ QCF as represented in \autoref{fig:concept}b}. Then, for any $I,J \in \mathbb{N}$ such that $I K \leq d_1-k$ and $J K \leq d_2-k$, the state $\ket{\Tilde{x}}$ produced after applying the convolution on the initial state $\ket{x}$ is such that:
\begin{equation}\label{eq:Convolution_example_2D}
    \sum_{i=I}^{I+K} \sum_{j=J}^{J+K} \Tilde{x}_{i,j} \ket{e_{d_1,i}, e_{d_2,j}} = U_{\textrm{Filter}}(\Theta) \sum_{i=I}^{I+K} \sum_{j=J}^{J+K} x_{i,j} \ket{e_{d_1,i}, e_{d_2,j}} \, \textrm{,}
\end{equation}
where $\Theta$ indicates the set of variational parameters associated to this layer, $\ket{e_{d_1,i}}$ is a single particle Fock state corresponding to \autoref{eq:Tensor_Encoding_modes}, and $U_{\textrm{Filter}}(\Theta) = (u_{i,j}(\Theta))_{i,j \in [K^2]}$ the QFC. The final state corresponds to a new tensor $\Tilde{x} = \left(\Tilde{x}_{i,j}\right)_{(i,j) \in [d_1]\times[d_2]}$ which is still tensor encoded. 

\subsection{State Injection based Pooling Layer}\label{subsec:SI_Pooling}

Pooling layers play a significant role in the CCNN architectures, as they allow one to reduce the dimension of the data through the computation. Usually, such a layer is followed by a nonlinear activation function \cite{montúfar2014numberlinearregionsdeep, raghu2017expressivepowerdeepneural}.
Here, we introduce a pooling layer that preserves the structure of the tensor encoded data provided in \autoref{eq:Tensor_Encoding_modes}, while reducing its size and applying a nonlinearity. The method presented here is based on state injection (see \autoref{fig:concept}b), a measurement-based technique that is suitable for near-term linear optical platforms, introduced in \cite{monbroussou2024quantum}.
Considering a tensor encoded input state, the pooling method consists of measuring half of the modes for each register. If a photon is measured in one mode, another photon is injected into the following one.
Because of the tensor encoding structure of the state, only one photon could be measured (and injected) per register, which leads to a low number of additional photons needed.
An illustration of this pooling layer is given in \autoref{fig:concept}b. The pooling operation performed is equivalent to the one in \cite{monbroussou2024subspace} when considering tensor encoding on Fock states instead of tensor encoding on states of fixed Hamming weight. The depth of this layer is $\mathcal{O}(1)$, and requires $k$ additional particles, by considering an adaptation of the State Injection. 
This pooling layer allows a reduction by half the size of the input state on each dimension, corresponding to the registers where half of the modes are measured. This operation is structurally similar to the classical average pooling, as explained in \autoref{chap:SI_Pooling}, and one could choose to measure a different number of modes to change the size of the pooling, as long as the structure of the output corresponds to the tensor encoding structure which allows to apply new convolutional layers to create complex neural network architectures.

\begin{figure*}[t]
     \centering
    \includegraphics[width=\textwidth]{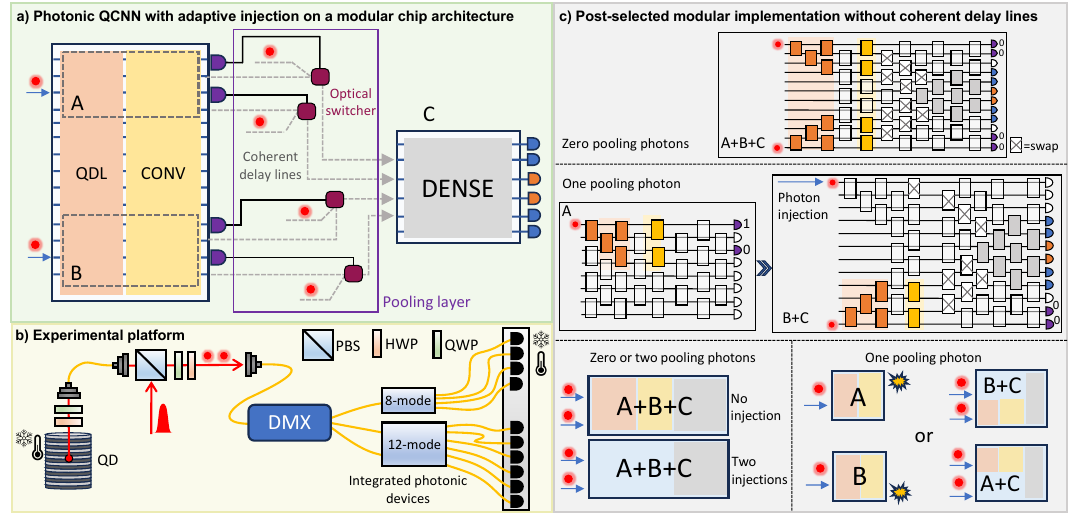}
    \caption{\textbf{QCNN architecture and experimental setup. a)} Modular architecture of a photonic QCNN with linear optics and adaptive state injection. The architecture consists of two linear optical circuits interleaved with a feedforward system. The first linear optical circuit includes the QDL, and the convolutional layer.
    The pooling layer, instead, requires a feedforward system through coherent delay lines to transfer the state after the convolutional into the dense layer, equipped with a fast optical switcher driven by the detectors of the pooling layer. The optical switcher injects (or not) a new single photon, or the state leaving the convolutional layer, according to the measurement outcomes of the detectors of the pooling layer. We experimentally encoded $4 \times 4$ images with rows and columns encoded into two separate registers, indicated here as $A$ and $B$, while  the dense layer is labeled as $C$. \textbf{b)} Scheme of the photonic hardware QOLOSSUS-2. Single photons are generated through a semiconductor quantum dot source (QD). They are then coupled into a temporal-to-spatial demultiplexer (DMX) and manipulated into an 8-mode and a 12-mode universal and programmable integrated interferometers. The output photon statistics are measured through superconducting nanowire single photon detectors (SNSPD). In the figure, we show only half of the detectors used, as we analyze all output modes for each integrated device. \textbf{c)} To cope with the absence of coherent delay lines and a feedforward system, the current implementation employs post-selection and a different separation of the stages $A,~B,~C$. In detail, above in the panel, we report the case in which no photons are detected in the pooling modes and, hence, the whole quantum state after the convolutional layer is injected into the dense layer through suitable swap operations. Parts $A,~B,~C$ are therefore realized jointly in the 12-mode circuit.
    In the middle section, we report the case in which a photon is found in the first pooling mode and, hence, a photon is directly injected into the dense layer for the first register. Here, the first register encoding ($A$) is performed in the 8-mode circuit, while parts $B,~C$ are jointly performed in the 12-mode one. All other cases are summarized in the bottom part of the panel.}
    \label{fig:exp_setup}
\end{figure*}

\subsection{Dense Layer}\label{subsec:Dense_Layer}

As in classical deep-learning architectures, the convolutional and pooling layers introduced previously are used in complex neural networks to extract important features and to reduce the size of the data during the computation. CCNN architectures are usually completed with a final dense layer, a linear neural network that concentrates most of the trained parameters and that extracts the key features to perform the learning task. For the present QCNN architecture, we propose to use a linear optical layer while merging each register from the tensor encoding structure. This circuit is a linear layer, indeed the following relation holds:
\begin{equation}\label{eq:Photonic_Dense_Linear}
    \rho_{\text{out}} = W(\Theta_{\text{Dense}}) \rho_{\text{in}} W^{\dagger}(\Theta_{\text{Dense}}) \, ,
\end{equation}
with $W$ the $\binom{m+k-1}{k} \times \binom{m+k-1}{k}$ unitary matrix corresponding to the action of the $m$-mode circuit in the subspace of $k$ photons, $\Theta_{\text{Dense}}$ the set of variational parameters, and $\rho_{\text{in}}$ the initial state. As explained in \cite{aaronson2011computational}, such Bosonic circuits are limited in their controllability, i.e., in the maximal number of free independent parameters. One could use state injection \cite{monbroussou2024quantum} layers to increase the Bosonic limit of $m^2-1$ parameters for a $m$ linear optical circuit while preserving the number of particles during the computation to avoid Barren plateaus. {We choose to focus on a linear optical dense layer in our experimental proposal to propose an architecture that can be verified in the very near term with a minimal number of adaptivity layers.}

The output of the QCNN architecture is a probability distribution obtained from the final measurement of the selected optical modes of the dense layer. Depending on the number of classes and the available detector technology, different measurement procedures can be employed to complete the architecture. 
Considering a typical classification task among a restricted number of $d$ classes, each label is assigned according to the probability to detect one or more photons in $d$ distinct modes or in $d$ bins of grouped modes. The training of the variational parameters of the dense layer according to a mean squared error loss ensures that the output label will be the correct one for the classification task. Further training procedure can be performed on the assignment of the labels to the output multi-photon configurations in the dense layer. This last procedure is investigated mainly in the experimental implementation as an additional readout layer that operates as a post-processing stage of the collected data. 

\subsection{Modular adaptive photonic architecture for PQCNN}
\label{subsec:ModularAdaptive}
Here, we discuss the overall architecture for the realization of the PQCNN modules described in the previous sections and sketched in \autoref{fig:exp_setup}a through an illustrative example. The adaptive photonic platform comprises two linear optical circuits interleaved with a feedforward system. The first linear optical circuit includes a QDL, where data is encoded in the amplitude of a quantum state, and then convolutional layer (CONV). These first two parts, named as $A$ and $B$ and identifying the registers employed to encode the images column and rows, can be realized within a single multi-mode linear interferometer. Such an optical circuit can be integrated into a miniaturized chip equipped with full programmability and control over all the internal parameters \cite{Carolan2015,Smith2022,maring2024versatile,Pentangelo2024, giordani2023experimental}. The current technology of integrated devices does not allow for rapid on-chip reconfigurability, the key element required to implement the pooling layer based on adaptive state-injections. Hence as an alternative solution one can realize the pooling layer outside the chip. This distributed design could have its own advantages if one wishes to integrate privacy where data encoding is desired to be separated from the computing layer as we discuss later. The adaptive state injection requires the measurements of some output of the convolutional layer and the sequential adaptive injection of photons into the final dense layer (part $C$). Therefore, the feedforward system envisages off-chip delay lines that preserve the coherence of the multi-photon state among the paths to transfer it after the convolutional layer into the dense layer, equipped with fast optical switchers driven by the detectors of the pooling layer which allow the injection of a new single photon if the detector of the pooling clicks. Finally, the dense layer is again a linear optical circuit realized with a second programmable integrated device. The internal parameters of the dense layer are trained to assign labels to the images according to the detection of photons in certain output configurations, the measurement strategy described in the previous section.

\begin{figure*}[htb]
    \centering
    \includegraphics[width=\linewidth]{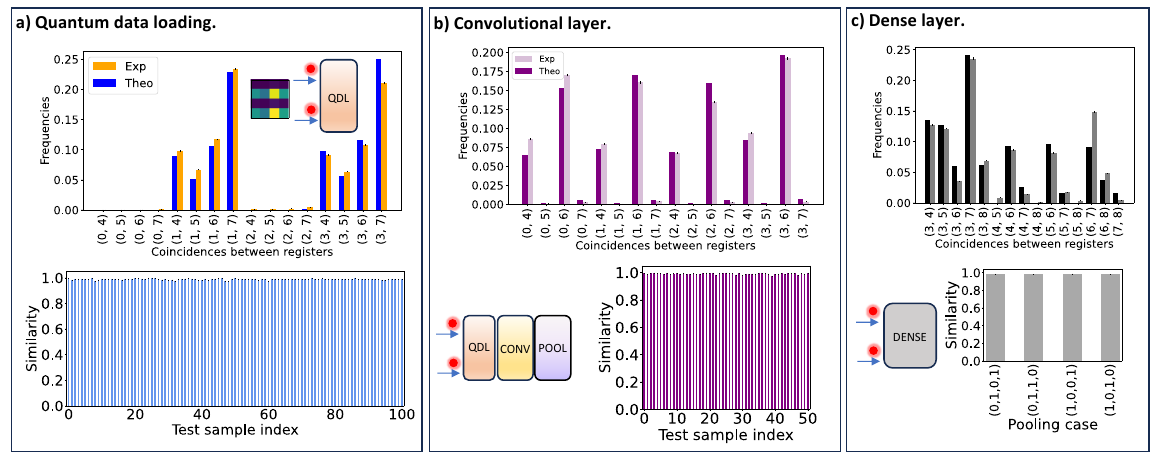}
    \caption{\textbf{Outcomes at each stage of the QCNN. a)} Above, the test image in the inset is encoded in the QDL stage.
    In the orange and blue bar plot, we show the theoretical and experimental probability distributions obtained at this stage for this particular image. 
    Below, the similarity between the theoretical and experimental distributions for hundred different images.
    \textbf{b)} The convolutional and pooling layers follow after the QDL stage. In the purple and lavender bar plot, we report the theoretical and experimental probability distributions at the output of the pooling layer for the image encoded in panel a). The purple bar plot shows the similarity between ideal distributions and experimental data obtained at the output of the pooling layer of fifty different images.  \textbf{c)} Here, the dense layer is tested separately from the previous layers. In the grey and black bar plot, we report the experimental results obtained for the unitary corresponding to the events where two photons are measured in the second and the last pooling modes, and two photons are injected directly into the dense layers, as shown in the inset. In the grey histogram, we show the similarity between theoretical and experimental output distributions for the four unitaries corresponding to such pooling cases.
    All uncertainties are estimated assuming Poissonian statistics.
    }
    \label{fig:results}
\end{figure*}

\section{Experimental apparatus}\label{sec:Experimental_Apparatus}

\textbf{Photonic hardware.} 
The previous model is experimentally tested with a hybrid quantum photonics platform, denominated QOLOSSUS-2, sketched in \autoref{fig:exp_setup}b, including two different integrated photonic devices with 8 and 12 modes, respectively. Single photons are produced through a commercially available (Quandela \emph{e-Delight}) semiconductor quantum dot (QD) single-photon source. It consists of an InGaAs matrix placed in a nanoscale electrically controlled micropillar cavity \cite{somaschi2016near} kept at cryogenic temperature ($\approx 4K$) through an \emph{Attocube-Attodry800} He-closed cycle cryostat. The QD is optically excited with a pulsed laser in resonance with the cavity characteristic wavelength ($928.05$~nm) \cite{somaschi2016near, loredo2019generation}. The repetition rate of the laser amounts to 160 MHz. The generated single photons are coupled into a single-mode fiber through a free-space confocal microscope mounted atop the cryostat shroud. Photons are then separated from the residual pumping laser in a cross-polarization scheme \cite{loredo2019generation} through the use of a polarizing BS and waveplates. A temporal-to-spatial demultiplexer (DMX) is employed to actively separate the stream of single photons. In particular, the DMX system exploits an acousto-optical modulator programmed to split the train of single photons into three spatially separated modes, which are then temporally synchronized via properly tuned in-fiber delay loops. 

After temporal synchronization, the multi-photon state is then injected into two different programmable universal integrated circuits with 8 and 12 modes respectively \cite{bellFurtherCompactifyingLinear2021,giordani2023experimental} fabricated through femtosecond laser {waveguide} writing \cite{Corrielli2021, Ceccarelli2020, Pentangelo2024}.
The on-device operations of the 12-mode device are controlled by thermo-optical phase shifters, through the application of external currents over the $132$ heaters on the top of the integrated device. In particular, the optical circuit was developed according to the universal design reported in \cite{bellFurtherCompactifyingLinear2021} in which Mach-Zehnder based configurations featuring a pair of reconfigurable internal phases enable the implementation of arbitrary unitary transformations within a more compact physical configuration. The 8-mode chip, designed according to the universal design of \cite{Clements:16} and encompassing 56 thermo-optic phase shifters, is employed to perform the QDL and convolutional layers only. 
The optical depth of the 12-mode chip is enough to allow for different internal configurations of the required building blocks i.e. QDL, convolutional, and dense layers. We note that the pooling layer is emulated in post selection due to the current unavailability of coherent delay lines and fast optical switches. This emulation procedure is described in the next paragraph. 
After the evolution within the integrated device, photons are detected with superconducting nanowire single-photon detectors (SNSPD) \cite{natarajan2012superconducting}.

\textbf{Implementation of the PQCNN via post-selection. }
In what follows, we tailor the description of the encoding of the PQCNN architecture to the experimental photonic platforms described above and reported in \autoref{fig:exp_setup}c.
The goal is to carry out a binary classification of $4 \times 4$ pixel images.
Firstly, we adapted the QDL layer to be encoded in the integrated photonic device that comprises 12 modes, which would in principle encode the full structure $A,~B$ (each comprising QDL and convolutional layer) $+C$ (dense layer). 
The size of the circuit limits the number of layers that can be reserved for the QDL. We opted for an experimental QDL composed of two 4-mode linear-optical circuits that use eight BSs of the device. Each circuit independently encodes a $4$-mode register (see \autoref{fig:exp_setup}c). As a result, the QDL produces tensor-encoded states capable of representing a subset of the possible $4\times 4$ pixel images, like grayscale bars and stripes. 
A suitable dataset was therefore adapted from the publicly available Pennylane Bars-and-Stripes (BAS) dataset, which is here denoted as Custom BAS dataset. Details about this dataset are provided in \autoref{chap:SI_qdl}.
The convolutional stage is encoded in one layer of the circuit with a total number of 4 tunable BSs, highlighted by the yellow area in \autoref{fig:exp_setup}c. The dense layer comprises 8 BSs of the 12-mode device.

The pooling is performed over four modes, two for each register. As previously said, the state injection is emulated via post-selection. Operationally, this means that different experiments are run with different numbers of injected photons and circuit configurations.
Some instances of the pooling configurations and the corresponding circuits are reported in \autoref{fig:exp_setup}c. The current implementation employs different separations of the stages $A,~B,~C$ into the 12- and 8-mode devices. In detail, the top panel of \autoref{fig:exp_setup}c reports the circuit configuration in the case in which no photons are detected in the pooling modes and, hence, the whole quantum state after the convolutional layer is injected into the dense layer through suitable swap operations. Parts $A,~B,~C$ are therefore realized jointly in the 12-mode circuit.
A second scenario is the case in which one photon is found in the first pooling mode in part $A$ and, hence, a photon is directly injected into the dense layer for the first register. Here, the first register encoding ($A$) is performed in the 8-mode circuit, while parts $B,~C$ are jointly performed in the 12-mode one (see middle panel of \autoref{fig:exp_setup}). Similar setup for the other scenario corresponding to one photon detected in the pooling mode of the second QDL register $B$. 

All the post-selection cases and the related circuits and measurement settings are briefly depicted in the last panel of \autoref{fig:exp_setup}c. To summarize, both the scenarios with zero photons in the pooling and no injections, and with two photons in the pooling and two injections, have been fully implemented in the 12-mode device by post-selecting on the output configurations that individuate each of the two configurations. The cases with one photon in the pooling layer are realized by encoding one QDL and one convolutional layer in the 8-mode device (part $A$ or $B$) and the second QDL, convolutional, and the dense layer in the 12-mode device.

\section{Results}

In this section, we present the results achieved with our experimental implementation of the previously described PQCNN architecture. 
At first, we present the experimental results obtained via the post-selection procedure illustrated in the previous section, 
demonstrating the feasibility of a near-term implementation of this algorithm. The results are presented in a step-by-step fashion, by analyzing the output at each layer to demonstrate full control over the platform. In particular, the sub-routines of QDL and convolutional are verified in both devices, the 8 and 12-modes, according to the $A,\,B, \, C$ scheme. 

Once the QDL and convolutional stages are demonstrated to agree with the theoretical expectations for both platforms, the post-selected output distributions of the PQCNN, which correspond to the various pooling cases, are combined according to the associated pooling probability, to carry out experimentally the binary classification task.

The last part of this section is dedicated to explaining how the current algorithm offers significant polynomial speed-ups compared to CCNN architectures by analyzing the scaling of the required quantum resources and its complexity .\\

\begin{figure*}
    \centering
    \includegraphics[width=\textwidth]{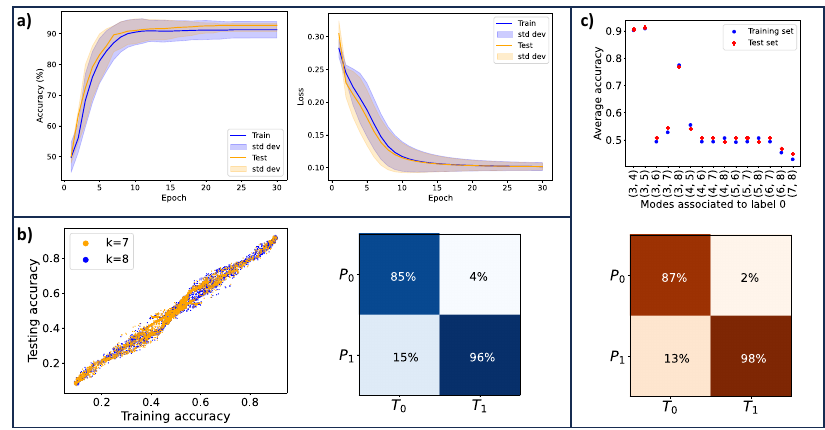}
    \caption{\textbf{Theoretical expectations and experimental results. a)} Accuracy and loss functions over 30 epochs during the offline training of the PQCNN. Training is performed over 400 images, while testing is performed over 200 images, all taken from the Custom BAS dataset.  \textbf{b)} Results of the experimental training of the readout layer. In the first strategy, all possible two-fold coincidence events are organized in clusters with seven and eight elements, associated with label 0. The events in the complementary clusters are associated with label 1. The training accuracy for the optimal clustering strategy over 400 samples amounts to $0.910$. On the right, we show the corresponding confusion matrix over 200 test samples. $P_{0/1}$ indicates predicted labels while $T_{0/1}$ stands for true label. The first column is normalized over the total number of Stripes images in the test set (101), while the second column over the total number of Bars images (99). The total accuracy over the test set amounts to $0.905$. \textbf{c)} Another employed strategy to train the readout layer, complementary to the previous one, consists of associating a group of modes with a label. In the upper plot, we report the performance of associating a given couple of modes to label 0. The best accuracy is achieved by associating output modes $(3,5)$ with label 0, with a training accuracy amounting to $0.909 \pm 0.008$ over 400 training samples and a testing accuracy amounting to $0.909 \pm 0.015$ over 200 testing samples, both averaged over 30 different reshufflings of the full dataset. Below, we show the confusion matrix for the same dataset considered in panel b), yielding a testing accuracy of $0.925$.}
    \label{fig:simulations_accuracy}
\end{figure*}

\subsection{Image binary classification via experimental PQCNN}
\noindent
\textbf{Preliminary numerical simulations.}
The structure of the PQCNN considered in the experiment for binary classification of $4\times4$ pixel images follows a preliminary study that identified the circuits of \autoref{fig:exp_setup}c as the optimal ones for encoding all the PQCNN layers and classifying the data. 
The performances of such a simulated PQCNN are summarized in \autoref{fig:results}a. Here, the training and the test stages of this network that follows the same structure as the experimental one, are reported. In particular, the optimization of the variational parameters of the convolutional and dense layer, namely the BS reflectivities, follows the procedure illustrated in Sections \ref{subsec:Conv_Layer}-\ref{subsec:Dense_Layer}. The trends of accuracy and loss functions are reported over 30 epochs of the training. The training set comprises 400 images, while the test set comprises 200 images. The images are taken from the Custom BAS dataset detailed in \autoref{chap:SI_qdl}.
The final training accuracy amounts to $91.3 \pm 2.6\%$ while the final testing accuracy to $92.7 \pm 2.1\%$, thus confirming the effectiveness of the architectures for the classification task under investigation. Numerical simulations on other datasets are discussed in the next section.

\noindent
\textbf{Experimental verification of intermediate layers.}
The first experimental verification consists of measuring the output single-photon distribution of the two QDL registers in the 8-mode and 12-mode devices. To this goal, we program the two optical circuits to implement solely the QDL layer, and measure the similarity with the expected distributions associated to each image. In particular, in the blue and orange bars of \autoref{fig:results}a, a single output distribution is reported as an example, corresponding to the image shown.
The experimental output distribution $p(x)$ is compared with the ideal one $q(x)$ by considering a similarity function $S$, defined as $S = \sum_x \sqrt{p(x) q(x)}$, where $x$ stands for an output configuration. The experimental average similarity over 600 images at this stage amounts to {$\bar{S} = 0.96907 \pm 0.00002 $} with the 12-mode circuit and to {$\bar{S} = 0.98946 \pm 0.00002 $} with the 8-mode circuit.

These results demonstrate that we are able to achieve full control over the two optical platforms to succesfully perform the QDL operation.

Once the 600 images have been encoded, they are fed to the convolutional and pooling layers.
Merging the QDL and CONV layers together requires five layers of BSs on the 8-mode and 12-mode chips and, hence, preliminary tests of these two sub-routines are performed on both devices. In \autoref{fig:results}b, the results obtained at the output of these stages are reported. {As a benchmark of the successful implementation of paired QDL and CONV layers, we obtain an average output similarity of {$\bar{S} = 0.94482 \pm 0.00003 $} with the 12-mode circuit and {$\bar{S} = 0.991923 \pm 0.000013 $} with the 8-mode circuit.

Finally, the dense layer is implemented in the 12-mode device (part $C$ of the network), and the output distribution is measured (see \autoref{fig:results}c). The settings considered for the BS reflectivities are the ones determined by the numerical simulation and training of the PQCNN. The dense layer can be tested by considering the two-photon injection configuration that happens when two photons are detected in the pooling modes and two new photons enter directly into the dense layer. The average similarity of this final stage is {$\bar{S} = 0.98486 \pm 0.00021 $}, averaged over the four possible unitaries.

All the measurements of such a preliminary characterization of the PQCNN stages have been performed by sending two photons into the devices and recording two-fold coincidences in the output.

\noindent
\textbf{Training and test of the experimental PQCNN.}
After such preliminary verification of correct loading of the images into quantum states and verification of the convolutional layer, the entire experiment is run 
with the addition of the dense layer. 

In the experimental verification of the PQCNN capability at classifying the Custom BAS dataset, we test a training strategy that can be effective in experimental conditions in which imperfections like incorrect optical circuit settings or bias in the detection efficiencies could hinder the performance of the network. The parameters of the dense layer have been set as follows.  
The reflectivities of the BSs are the optimal ones calculated according to the offline training stage previously run on the theoretical and simulated PQCNN.
The phases among the paths are set randomly instead. We use 400 of the 600 images as a training set to optimize the accuracy in the classification by finding the best binning of the dense layer's output modes.  More precisely, this further training procedure aims to test whether a convenient labeling of the dense layer's output modes can also be used as a further layer of the network, namely a readout layer. 
In particular, the accuracy after considering such an additional readout layer is maximized over the training set by varying the detector binning that predicts label 0 or 1 for each image. Two strategies have been pursued. In the first one, all possible two-mode configurations in which one can find two photons in the six output modes of the dense layer, that are in total 15, are organized in clusters with $k$ pairs associated with label 0 and $15-k$ pairs associated with label 1. For each possible clustering configuration for $k=7$ and $k=8$ we compute the resulting classification accuracy over the training set together with the corresponding test set accuracy as reported in \autoref{fig:simulations_accuracy}b. The best accuracy reached on the training set with the optimal clustering associations is $0.910$, and the one on the test set for the same clustering amounts to $0.905$. The second strategy to train the readout layer consists of associating the labels with a group of output modes. For example, a possible choice is to consider a pair of modes associated with the label 0. The training carried out in \autoref{fig:simulations_accuracy}c according to such an association shows that the best training accuracy is achieved by setting output modes $(3,5)$ with label 0, which amounts to $0.909 \pm 0.008$. The accuracy over the test set is $0.909 \pm 0.015$. Both quantities are averaged over 30 different reshufflings of the full dataset.}

\subsection{Running time and Simulations}\label{subsec:Simulations_Complexity}

The PQCNN architecture here proposed offers a polynomial advantage over CCNN architectures. In particular, the degree of polynomial advantage depends on the number of input photons, which is given by the dimension $k$ of the input tensor. In \autoref{table:Time_Complexity} we compare layer-by-layer the number of operations, and so the time complexity, required to implement a CCNN in terms of the parameters that individuate the size of the network, namely $K_i, \, d_i, \, k, \, m$ defined in Secs. \ref{subsec:Conv_Layer}- \ref{subsec:Dense_Layer},  {with the ones required by a PQCNN.}
The running time of the PQCNN is related to several features of the adopted platform: the characteristics of the optical apparatus, including coupling and {propagation} losses, brightness of the single-photon source{, detection efficiency, the number and reconfiguration speed of adaptive injection layers.}

\begin{table}[h!]
\centering
\begin{tabular}{ |c|c|c|c| } 
    \hline
    & Convolutional layer & Pooling layer & {Dense layer} \\ 
    \hline
    CCNN & $\mathcal{O}( \prod_{i=1}^k K_i^2 \cdot d_i^2)$ & $\mathcal{O}(\prod_{i=1}^k d_i)$  & $\mathcal{O}(\binom{m+k-1}{k}^2)$  \\ 
    \hline
    PQCNN & $\mathcal{O}(K)$ & $\mathcal{O}(1)$ & $\mathcal{O}(\frac{p}{m})$ \\
    \hline
\end{tabular}
\caption{Comparison between the  CCNN and PQCNN required resources. We consider $k$ dimensional convolutional neural network layers with $d_1 \times \cdots \times d_k$ the size of the square input tensor, $d = \max(\{d_i\}_{i=1}^k)$, $\{K_i\}_{i=1}^k$ the size of the convolutional filter, $K = \max(\{K_i\}_{i=1}^k)$, and $p$ the number of parameters in the dense layer. {We call $m$ the total number of modes with $m = \sum_{i=1}^k d_i$.}}
\label{table:Time_Complexity}
\end{table}

\autoref{table:Required_Ressources} summarizes the resources of each layer in the PQCNN, both for the adopted scheme and for the most general architecture which goes beyond the one illustrated in \autoref{fig:concept}, for what concerns the number of modes, input photons, and injected photons. In this Table, we consider for each layer the same input tensor size. In practice, the input tensor size could decrease during the computation because of the pooling layers, but one can choose to use additional modes or photons in a custom architecture. 
\begin{table}[h!]
\centering
\begin{tabular}{ |c|c|c|c|c| } 

    \hline
     \textbf{Adopted} & QDL & Convolutional layer & Pooling layer & Dense layer \\ 
    \hline
    Modes & $m$& $m$& $m$& $m/2+\alpha$ \\
    \hline
    Photons & $k$ & $k$ & $2k$ & $k$ \\
    \hline
    \hline
    \hline
    \textbf{General} & QDL & Convolutional layer & Pooling layer & Dense layer \\ 
    \hline
    Modes & $m'$ & $m'$ & $m'$ & $m'+\alpha'$ \\
    \hline
    Photons & $k'$ & $k'$ & $2k'$ & $k'$ \\
    \hline
\end{tabular}
\caption{Resources required for each layer to build the PQCNN setup. We consider convolutional neural network layers with $d_1 \times \cdots \times d_k$ the size of the square input tensor for each layer (with $m' = \sum_{i=1}^k d_i$ modes and $k'$ photons). We call $\alpha' \in \mathbb{N}$, the number of mode one can add to the dense layer. For the experimental architecture developed in this work, $m=8$, $k=2$, and $\alpha=2$.}
\label{table:Required_Ressources}
\end{table}
\begin{table}[h!]
\centering
\begin{tabular}{ |c|c|c|c|c| } 
    \hline 
    & Input Size & \# Parameters & Train & Test \\
    \hline
    \hline
    BAS & $4 \times 4$ & $10$ & $93.7 \pm 1.6 \%$ & $93.0 \pm 1.2 \%$ \\
    \hline 
    Custom BAS & $4 \times 4$ & $10$ & $91.3 \pm 2.6$ \% & $92.7 \pm 2.1$ \%\\
    \hline
    MNIST & $8 \times 8$ & $30$ & $95.1 \pm2.9 \%$ & $93.1 \pm 3.6 \%$ \\
    \hline
\end{tabular}
\caption{Classification Accuracy for the PQCNN architecture, for different datasets and architectures.}
\label{table:Simulation_Classification}
\end{table}
The number of pooling layers, which are the most challenging part from the hardware point of view, depends on the design choice of the architecture. It usually increases with the input size as pooling layers reduce the size of the input image while convolution layers extract the important features. However, as each pooling layer reduces by half the size of the input, the number of pooling layers will increase logarithmically with the problem size. Notice that one could easily adapt this layer to achieve a reduction of higher or lower order by simply measuring more or fewer modes.

In \autoref{table:Simulation_Classification}, we present simulation results for datasets of different sizes.
First, the experimental equivalent model is trained for the Custom BAS dataset described in \autoref{chap:Encoding}, i.e., the samples that can be encoded with the adopted experimental setups. Then we compare for the Pennylane Bars and Stripes (BAS) dataset with $4 \times 4$ images. 
Finally, we compare the architectures for the MNIST dataset \cite{LeCun_LeNet} made of $8 \times 8$ images, by considering a $m=16$ mode-circuit with $k=2$ initial photons, with a single convolutional, a single pooling layer requiring $2k=4$ photons in total, and no additional modes ($\alpha=0$) used in the dense layer. The results for those simulation are close to the results for fault-tolerant quantum architectures of QCNN \cite{Li_2020, kerenidis2019}. The details on each simulation, {and equivalent used models} can be found in \autoref{chap:Simulation_Results}.

{All the simulation{s} have been {performed} using our open source library that is {tailor-made} for QML photonic algorithms by performing the computation in the most {{suitable} subspaces. This Pytorch \cite{paszke2019pytorchimperativestylehighperformance} based toolkit could be of independent use for photonic simulation and can be found in \cite{PhotonicSubspaceQMLToolkit}}.

\section{Discussions}
This work investigates an architecture for QCNNs based on the properties of linear optical platforms. 
The introduced PQCNN establishes a direct relationship between the features of Hamming-weight preserving circuits, which provide advantages against the Barren plateaus often affecting the training of quantum machine learning protocols, and linear optical ones, equipping the latter with nonlinearities coming from the recently introduced photonic state injection technique.
The map between subspace preserving and photonic QCNNs is first theoretically described layer-by-layer, and then demonstrated experimentally relying on 8-mode and 12-mode universal reconfigurable integrated photonic devices, where adaptive state injection is emulated through post-selection procedures.
The demonstration involves a binary classification of 4-by-4 pixel images taken from the public repository \textit{Pennylane Bar-and-Stripes} and tailored to the characteristics of the employed platform. The outcome of each layer is first analyzed individually and, afterward, the full PQCNN is implemented. Finally, 
the architecture proposed here is analyzed in terms of the quantum resources like the number of photons, pooling layers, and the optical circuits depth required for its implementation. Such a study highlights a good scaling of the architecture with promising applications in larger-scale systems. 

In this direction, future implementation would leverage genuine adaptive state injections, realized through coherent connections among the various PQCNN modules and fast switches in between, activated by fast single-photon detectors. The realization of such a photonic hardware implies facing some technological challenges, like the realization of coherent and stable links among different multiport interferometers and fast reconfigurability. 
Among the possible near-term solutions for coherent transmissions of photonic encoded states, we foresee the use of chip-to-chip connections based on polarization encoding that have been recently investigated for sharing quantum states among integrated devices through path-to-polarization converters \cite{Wang:16}. While on-chip fast operations remain challenging due to the limited speed of thermal heaters \cite{wang2020integrated, Giordani2023_rev} or to the high losses of active materials and components \cite{thomas2024noise, memeo2024micro, Zhang2021}, the current technology of bulk and in-fiber optical switches allows for operations in the GHz range with limited photon losses, thus further motivating investigations oriented towards off-chip realizations of adaptive state injections. Following such an approach, the development of fast detectors would allow for near term implementations of genuine adaptive photon injections on the $ns$ time scale \cite{Psi_Quantum}. 
Furthermore, the intrinsic distributed nature of the PQCNN naturally aligns with recent advances in privacy-preserving quantum computing. In particular, the recently published works on distributed blind quantum computing \cite{Polacchi2023,Polacchi2024} enable secure, distributed computation without revealing sensitive information, making it fully compatible with the current PQCNN architecture.

In summary, the theoretical and experimental results here reported provide a further perspective picture of adaptivity, in the form of photonic state injection, as an additional nonlinear ingredient for linear optics-based quantum machine learning protocols, thus triggering {the investigation of new algorithms} inspired by and tailored to quantum photonic platforms. 

\section*{Acknowledgements}
The authors acknowledge the support of the European Union’s Horizon Europe research and innovation program under EPIQUE Project (Grant Agreement No. 101135288), the ERC Advanced Grant QU-BOSS (QUantum advantage via nonlinear BOSon Sampling, grant agreement no. 884676), and the ICSC--Centro Nazionale di Ricerca in High Performance Computing, Big Data and Quantum Computing, funded by European Union--NextGenerationEU. EK acknowledge support from the EPSRC Quantum Advantage Pathfinder research program within the UK’s National Quantum Computing Center.

\bibliography{biblio.bib}

\onecolumngrid
\newpage
\appendix
\input{Supplementary_Materials/Appendix_RBS_BS}
\input{Supplementary_Materials/Appendix_Encoding}
\input{Supplementary_Materials/Appendix_Conv_Layer}
\input{Supplementary_Materials/Appendix_Pooling}
\input{Supplementary_Materials/Appendix_Simulations}
\input{Supplementary_Materials/Details_setup}

\end{document}

%% file: Supplementary_Materials/Appendix_RBS_BS.tex
\section{Connection between Reconfigurable Beam Splitters and photonic Beam Splitters}\label{chap:RBS_Photonic_BS}

\begin{figure}[h!]
\includegraphics[height=0.31\textwidth]{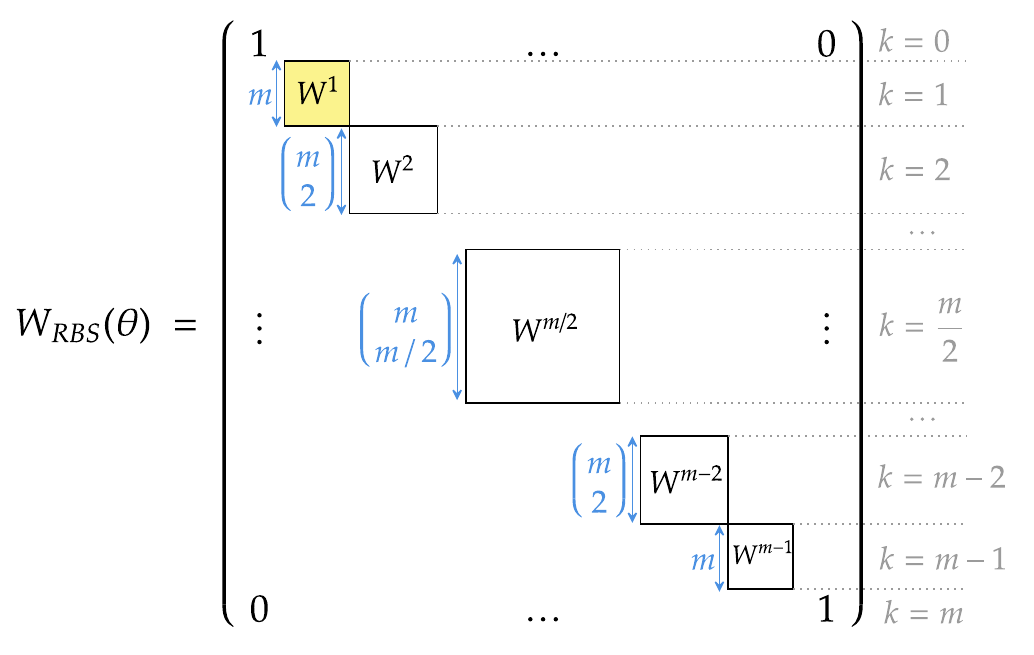}
\includegraphics[height=0.31\textwidth]{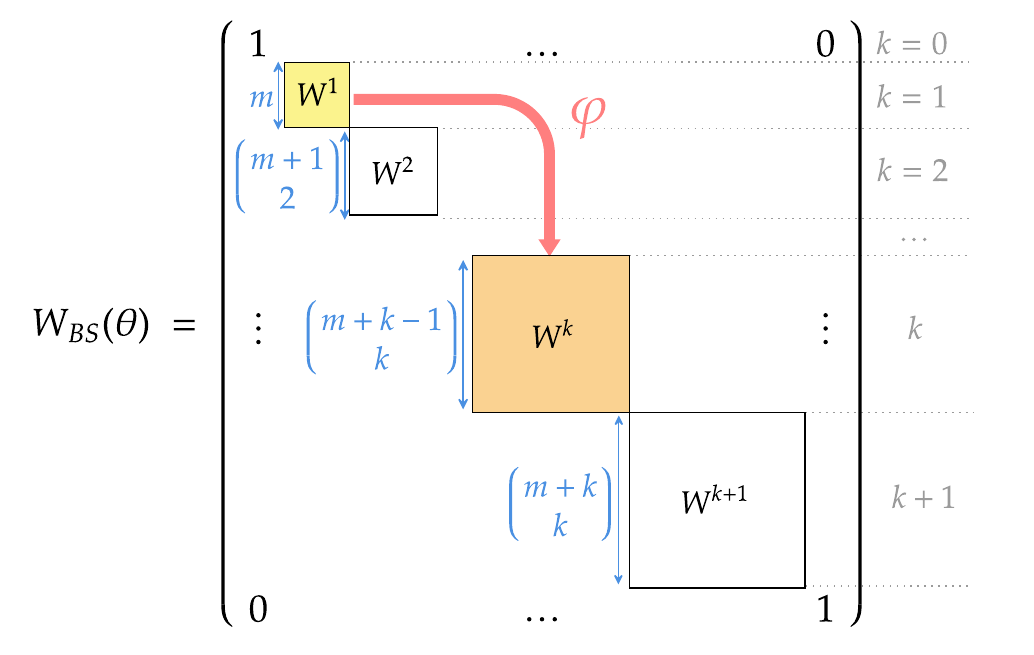}
\caption{Representation of RBS based (left) and of linear optic (right) quantum circuit equivalent unitary matrices. Each bloc $W^k$ corresponds to a subspace of HW $k$ (left) or to a subspace of $k$ particles (right). $\varphi$ corresponds to the photonic homomorphism that connect each subspace of fixed number of particles.}
\label{fig:RBS_BS_Bloc_Unitary}
\end{figure}

To perform the quantum convolutional layer in \cite{monbroussou2024subspace}, one need to use the Reconfigurable Beam Splitter (RBS) gate. This gate is widely used for HW preserving algorithms \cite{Landman2022, Cherrat2024quantumvision, kerenidis2022quantum, jain2023quantumfouriernetworkssolving, coyle2024, raj2025}. The Reconfigurable Beam Splitter (RBS) gate is a 2-qubit gate that corresponds to a $\theta$-planar rotation between the states $\ket{01}$ and $\ket{10}$:
\begin{equation}\label{eq:RBS_2_qubit_gate}
    W_{RBS}(\theta) = \begin{pmatrix}
        1 & 0 & 0 & 0 \\
        0 & \cos(\theta) & \sin(\theta) & 0 \\
        0 & -\sin(\theta) & \cos(\theta) & 0 \\
        0 & 0 & 0 & 1 \\
        \end{pmatrix} \, \textrm{.}
    \end{equation}    
One of the main elements of Linear Optical circuits is the Beam Splitter (BS), which is a two modes gate that acts on the amplitude of input photons. It is well known that lossless two modes BS (with two input and output modes) in quantum optics is described by the unitary matrix $W_{BS}$ which has the form~\cite{makarov2022theory}:
\begin{equation}
    \binom{\hat{b}_1}{\hat{b}_2} = W_{BS}(T,R,\phi) \binom{\hat{a}_1}{\hat{a}_2}, \quad W_{BS}(T,R,\phi) = \begin{pmatrix} \sqrt{T} & e^{i\phi} \sqrt{R} \\
    -e^{-i\phi}\sqrt{R} & \sqrt{T} \end{pmatrix} \, ,
\end{equation}
with $\hat{b}_1$, $\hat{b}_2$ the ouptut mode annihilation operators, $\hat{a}_1$, $\hat{a}_2$ the input mode annihilation operators, $T$ and $R$ are the transmittance and reflectance ($R+T=1$), and $\phi$ is the phase shift. In this work, we will focus on the ideal model with the phase shift $\phi=0$. Using a change of variable, it comes that:
\begin{equation}
    W_{BS}(\theta)= \begin{pmatrix} \cos(\theta) & \sin(\theta) \\
    -\sin(\theta) & \cos(\theta)\\
    \end{pmatrix} \quad \quad \text{with} \quad \cos(\theta) = \sqrt{T} \quad \text{and} \quad \sin(\theta) = \sqrt{R} \,\textrm{.}
\end{equation}
Both BS and RBS perform a $\theta$-planar rotation between state $\ket{01}$ and $\ket{10}$, but those states are Fock states in the photonic case, and qubit states in the HW preserving case. However, BS and RBS are subspace-preserving gates as RBS gates preserve the HW and BS gates preserve the number of particles. As a result, one can express the equivalent unitary of both BS and RBS based quantum circuits as block-diagonal as explained in \cite{monbroussou2024trainabilityexpressivityhammingweightpreserving} and  \cite{monbroussou2024quantum}. We illustrate those block diagonal equivalent unitary matrices in \autoref{fig:RBS_BS_Bloc_Unitary} for $m$ qubits and $m$ modes. In these figures, each block represents the equivalent unitary when considering a fixed HW $k$ or a fixed number of particles $k$. For the photonic case, the number of particles is unbounded, and the equivalent unitary is of infinite dimension. Both gates have the same impact on the initial state $\ket{00}$, $\ket{10}$, and $\ket{01}$ but they act differently on the state $\ket{11}$. In addition, other initial Fock state can be considered for the photonic BS that does not map to any state for the RBS. 

Consider two quantum circuits, the first is a photonic circuit of $m$ modes, and the second is a $m$-qubit quantum circuit. When considering the subspace of HW $1$ (unitary subspace) for the RBS and the subspace of a single particle for the BS, a BS applied between modes $i$ and $j$ in the first circuit, and a RBS applied between qubit $i$ and $j$ will have the same effect. This is the reason why one can easily adapt the convolutional layer presented in \cite{monbroussou2024subspace}, as the tensor encoding described in Eq.~(1) of the main text ensures that, for each register, there is only one particle. 

For larger subspaces, the equivalent unitary matrices will differ. First, the size of the subspace for $m$ qubits and HW $k$ corresponds to the number of bitstring of $m$ bits and HW $k$ and is $\binom{m}{k}$, while the size of the subspace for $m$ modes and $k$ particles is $\binom{m+k-1}{k}$. In addition, authors in \cite{aaronson2011computational} explain that the homomorphism $\varphi$ describes the way in which second-quantization enforces the evolution of $k$ indistinguishable bosons, given the unitary evolution for a single boson. This relationship and its impact on linear optical quantum circuit controllability are discussed in \cite{monbroussou2024quantum}. For HW preserving quantum circuits, a similar relationship between the block can exists. For example, when considering only a line connectivity, RBS acts as a Fermionic BeamSplitter gates and each block $W^k$ is the $k$-compound matrix of $W^1$ \cite{kerenidis2022quantum}. For a greater connectivity, such relationship disapears and even if all the blocs are highly correlated, each subspace can be perfectly controled and are not always determined by the first one \cite{monbroussou2024trainabilityexpressivityhammingweightpreserving}.

%% file: Supplementary_Materials/Appendix_Encoding.tex
\section{Quantum Data Loading}\label{chap:Encoding}
\label{chap:SI_qdl}
In this Section, we discuss the Quantum Data Loading part of the architecture. As explained in \autoref{subsec:Tensor_Encoding}, the Photonic QCCN is based on the tensor encoding. We recall the expression of the corresponding state for a classical tensor of dimension $k$ such that $x = (x_{1, \dots, 1}, \dots,  x_{d_1, \dots, d_k}) \in \mathbb{R}^{d_1 \times \dots \times d_k}$. The corresponding photonic tensor encoded state is described by \autoref{eq:Tensor_Encoding_modes}:
    \begin{equation*}
        \ket{x} = \frac{1}{||x||} \sum_{i_1 \in [d_1]} \dots \sum_{i_k \in [d_k]}  x_{i_1, \dots, i_k} \ket{e_{d_1, i_1}} \otimes \dots \otimes \ket{e_{d_k, i_k}} ,
    \end{equation*}

where $\ket{e_{d_l, i_l}} = \ket{0 \dots 0 1 0 \dots 0}$ represents a Fock state over $d_l$ modes, with a single excitation (photon) in mode $i_l$ and vacuum in all other modes. To encode a normalized tensor of size $d_1 \times d_2 \times \dots \times d_k$ from an input Fock state of $k$ particles, one need to use a quantum circuit with at least $\prod_{i=1}^k d_i -1$ degrees of freedom, i.e., a quantum circuit that can freely control the amplitudes of the output state in the following basis:
\begin{equation}\label{eq:Basis_Tensor_Encoding}
    B = \left\{ \ket{e_{d_1, i_1}} \otimes \dots \otimes \ket{e_{d_k, i_k}} \right\}_{(i_1, \dots, i_k) \in [d_1] \times \dots \times [d_k]}
\end{equation}

However, a linear optical circuit is limited in its controlability for input states with several particles as explained in \cite{aaronson2011computational} due to the photonic homomorphism illustrated in \autoref{fig:RBS_BS_Bloc_Unitary}. In the following, we first explain how we encode our data for the experiment introduced in \autoref{sec:Experimental_Apparatus}. Then we propose possible way to encode larger data points on a larger photonic architecture.

\subsection{Quantum Data Loading in the Experiment considered}

\begin{figure}[h!]
    \centering
    \includegraphics[width=0.95\linewidth]{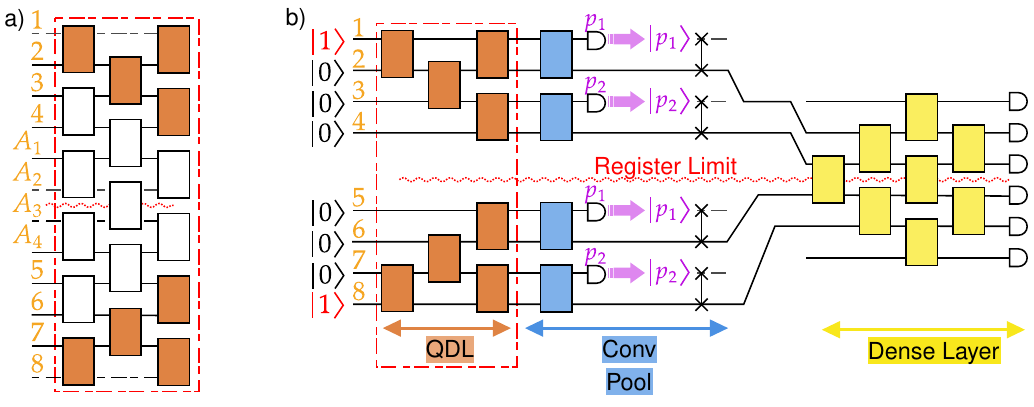}
    \caption{\textbf{Photonic QCNN architecture for experimentation. a)} Part of the $12 \times 12$ mode circuit used to encode the classical data. \textbf{b)} Photonic QCNN architecture experimentally tested. The red dotted curvy line represent the separation between the line and column registers.}
    \label{fig:QCNN_eq_QDL}
\end{figure}

The Photonic QCNN algorithm is experimentally tested with a hybrid quantum photonic platform sketched in Fig.~\ref{fig:exp_setup}a. A quantum data-loader, a quantum convolutional layer, a pooling layer, and a final dense layer are created using the $12$-mode programmable integrated interferometer as described in Fig.~\ref{fig:exp_setup}c. The Photonic QCNN architecture chosen is represented in Fig.~\ref{fig:QCNN_eq_QDL}b, and the 

\begin{figure}
    \centering
    \includegraphics[width=0.25\linewidth]{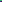}
    \hspace*{0.2in}
    \includegraphics[width=0.25\linewidth]{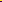}
    \hspace*{0.2in}
    \includegraphics[width=0.25\linewidth]{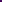}
    \caption{From left to right: sample from BAS dataset; plain sample with all pixels with the same value; sample from the Custom dataset, performed by applying a Gaussian noise on the plain sample corresponding QDL parameters.}
    \label{fig:dataset_sample_examples}
\end{figure}

Due to the limitation of the chip size, we encode our data only using $3$ parameters per register, without any parameter that link the registers. As a result, we are constrained in the amount of classical data that can be encoded. For example, the Pennylane Bars and Stripes (BAS) dataset required to encode any sample $x \in \mathbb{R}^{4 \times 4}$ which is not possible with our architecture. Therefore, we design the Custom BAS dataset in order to only consider samples of size $4 \times 4$ pixels but with more structure in order to allow our experimental linear optical data loading circuit to work. To do so, we simply choose to first design a test dataset made of bars and plot image, but with all the bright pixels to be equal in value, and all the dark pixels to be equal to $0$. Such image is easy to load considering our experimental QDL (see \autoref{fig:QCNN_eq_QDL}): for an image with lines, one just needs to tune the BS on the line register to have the photon in a uniform superposition on the corresponding modes, and the second photon on the column register to be uniformly distributed. Then we design a Custom dataset by applying a Gaussian noise on the corresponding QDL parameters for each possible set of lines or bars. An illustration is given in \autoref{fig:dataset_sample_examples}.

\subsection{Quantum Data Loading for large data}

For complex learning problem, the QCNN architecture must use a QDL that can perform the tensor encoding of large tensors. To do so, a QDL needs enough "controllability", meaning that it must be able to freely control the amplitudes of the orthogonal state basis described in \autoref{eq:Basis_Tensor_Encoding}. We refer to this number of controllable orthogonal states as \textit{degrees of freedom}. The limitation of $m$-mode linear-optical circuit controlability is explained in \cite{aaronson2011computational} where the photonic homomorphism (see \autoref{fig:RBS_BS_Bloc_Unitary}) limits the number of degrees of freedom to $m^2-1$, or $m(m-1)/2$ if we are not considering the phases of the state as in the tensor encoding. We can suggest several solutions in order to increase the controllability of a photonic QDL. First, one can consider additional ancillas mode to increase the controlability. On can go beyond the photonic homomorphism limitations by considering non-linearities, post-processing, or adaptivity scheme \cite{chabaud2021quantum, monbroussou2024quantum}.

%% file: Supplementary_Materials/Appendix_Conv_Layer.tex
\section{Convolutional Layer}\label{chap:Appendix_Convolutional_Layer}

In this Section, we remind how the Hamming-weight (HW) convolutional layer from \cite{monbroussou2024subspace} works, and how it can be adapted to linear optics. First, we recall the definition of a classical convolutional layer. Consider an input tensor $x \in \mathbb{R}^{d_1 \times \dots \times d_k}$ and a $D$-dimensional convolutional layer applied the first $D$ dimensions of the input, with a filter of size $K_1 \times \dots \times K_D$. The convolutional layer 
derives a tensor $\Tilde{x} \in \mathbb{R}^{\Tilde{d}_1 \times \dots \times \Tilde{d}_k}$, with $\Tilde{d}_i = d_i - 2 \lfloor \frac{K_i}{2} \rfloor$ if $i \leq D$ and $\Tilde{d_i} = d_i$ else, such that:
\begin{equation}\label{eq:Classical_Convolution_Layer}
    \Tilde{x}_{i_1, \dots, i_k} = \sum_{k_1 = 1}^{K_1} \dots \sum_{k_D = 1}^{K_D} W_{k_1, \dots, k_D} x_{i_1 -\lfloor \frac{K_1}{2} \rfloor +k_1, \dots,  i_D -\lfloor \frac{K_D}{2} \rfloor +k_D, i_{D+1}, \dots, i_k}
\end{equation}
with $W$ the convolutional filter. Notice that this operation corresponds to a discrete convolution that is invariant per translation.
The output tensor dimensionality typically differs from the initial input one, and can be chosen by employing padding techniques, i.e., by considering an extended initial input. The additional parts of the input tensor due to padding are usually filled with $0$, or a repetition of the edge values.

\begin{figure}[h!]
\centering
    \includegraphics[height=0.25\textwidth]{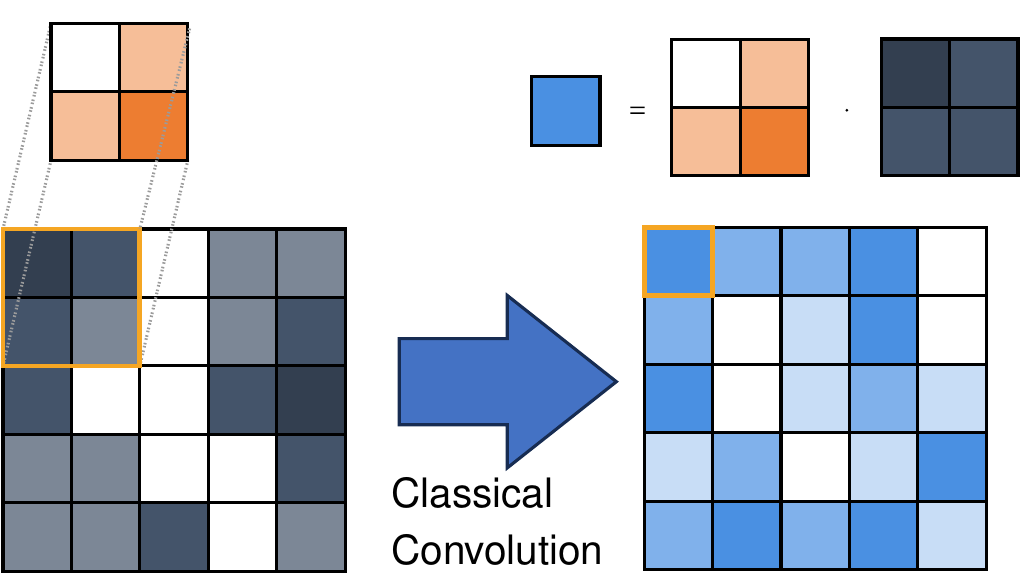}
    \hspace*{0.2in}
    \includegraphics[height=0.25\textwidth]{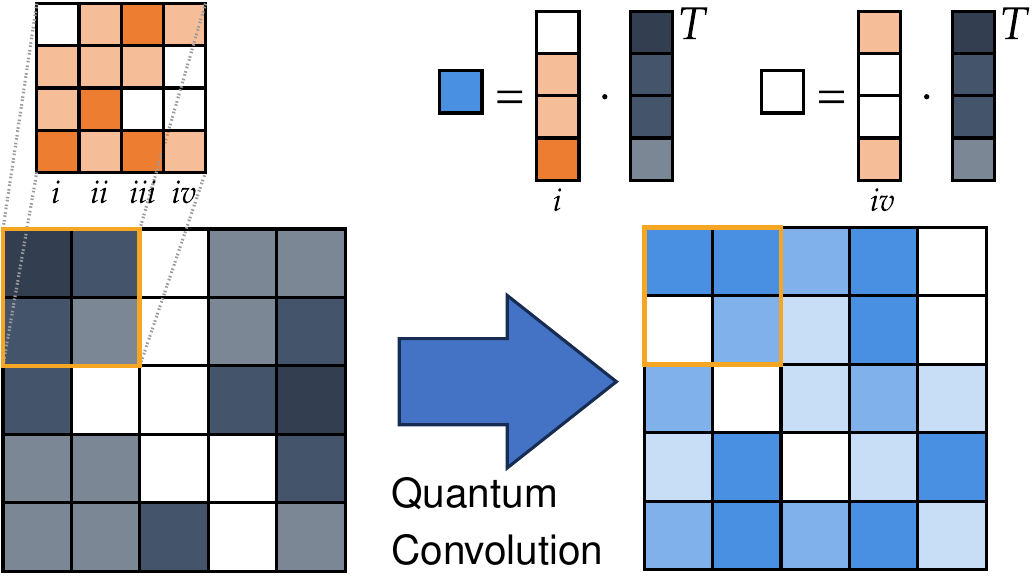}
\caption{Classical (left) and quantum (right) convolutional layer illustration. For both examples, the input tensor is the same $2$-dimensional image, and both filters are of the same size. The orange square on the input image represents the convolution window considered for the described pixel evolution.}
\label{fig:Classical_VS_Conv_Layer}
\end{figure}

For the quantum convolutional layer described in \autoref{subsec:Conv_Layer} of the main text, the input tensor is a quantum state $\ket{x}$ that is tensor encoded as given in Eq.~(1) of the main text. The output state $\ket{\Tilde{x}}$ is still tensor encoded, and the layer preserves the dimension and sizes of the initial state. The quantum convolutional layers also performs a similar operation to \autoref{eq:Classical_Convolution_Layer}, as each new coefficients of $\Tilde{x} \in \mathbb{R}^{d_1 \times \dots \times d_k}$ is a linear combination of the pixel amplitudes in its neighborhood. The main difference between the quantum and the classical convolutional layer is that this notion "neighborhood" differs. For the classical layer, the neighborhood, is defined as the closest pixels in a convolution window given by the size of the filter, and the linear combination is always the same as defined in \autoref{eq:Classical_Convolution_Layer}. For the quantum layer, the convolution window is defined by the position of the circuits in each register. The neighborhood is the same, and each linear combination is defined through a column of the equivalent orthogonal matrix that represents the action of each circuits on the subspace of affected state. An illustration of both operation is given in \autoref{fig:Classical_VS_Conv_Layer}. As a result, the quantum convolutional layer is more structured, as the filter is defined through a $(\prod_{i=1}^D K_i) \times (\prod_{i=1}^D K_i)$ orthogonal matrix, with a maximal of $\sum_{i=1}^D K_i(K_i -1)/2$ free independent parameters that corresponds to the maximal controlability of each circuit in the subspace of 1 particle. For the classical layer, the convolutional filter is a $D$-dimensional tensor of size $K_1 \times \dots \times K_D$, and the maximal number of free independent parameters is thus $\prod_{i=1}^D K_i$.

In deep-learning architecture, one could choose to tune the stride of the convolutional layer, i.e., the size of the step for the convolutional window. Increasing the stride further reduces the dimension of the output image, which makes this hyperparameter an integer factor of downsampling. One could argue that the quantum convolutional layer behaves like a convolutional layer with the stride equal to the filter size on each dimension, as all convolution windows are separated by these distances. One could ensure that each pixel in the input tensor is involved in a convolution operation with a specific convolution window by loading a shifted version of the input on an additional tensor-dimension, by increasing the number of particles and adding modes in a new register.  

%% file: Supplementary_Materials/Appendix_Pooling.tex
\section{State Injection based Pooling Layer}\label{chap:SI_Pooling}

In this Section, we recall the definition of State Injection and how to use it in the photonic quantum convolutional architecture to produce a Pooling layer. We highlight the similarities of the proposed Pooling layer with the usual combination of Average Pooling and activation functions used in classical deep learning architectures.

\begin{definition}[State Injection from \cite{monbroussou2024quantum}]\label{def:StateInjection}
    We call \textbf{State Injection} (SI) any operation on an $m$-mode photonic platform that performs photon-counting measurements in one or several modes, and, depending on the outcomes obtained, re-injects some photons back in one or several modes.
    Overall (since no single outcome is post-selected on), this operation is described by a CPTP map on the relevant Hilbert space.
    Different SI operations hence correspond to different choices of modes that undergo measurements and/or re-injections, and different choices of rules that map measurement outcomes to the corresponding re-injections that should be performed. We refer to the latter as a choice of \textbf{injection functions}.
\end{definition}

\begin{figure}[h!]
    \centering
    \includegraphics[width=1.0\linewidth]{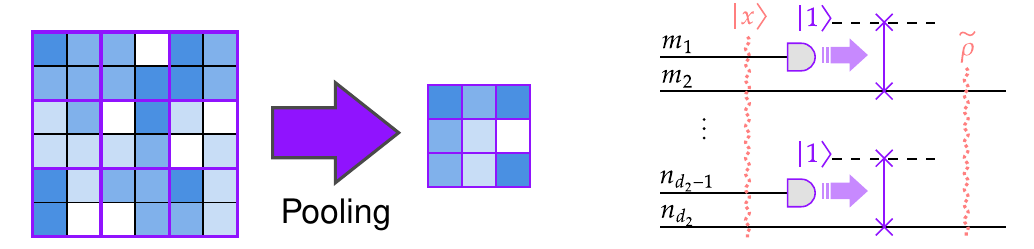}
    \caption{State Injection based Pooling Layer for photonic QCNN architectures for a $2$-dimensional tensor input. In purple, we illustrate the Pooling windows of each coefficient for the output tensor.}
    \label{fig:SI_Pooling}
\end{figure}

The photonic pooling layer introduced in Sec.~IIC of the main text consists of measuring every two modes of each register on which one desire to reduce the size by half. Considering the tensor encoding described in Eq.~(1) of the main text, if one photon is measured in a mode, there is no more particles in the corresponding register. We use SI to inject a new photon on the following mode, ensuring that the output state in the remaining modes of each register conserve the tensor encoding structure (with a single particle per register). Notice that, by considering this Pooling method, each new coefficient in the output tensor is defined only according to the value of previously neighbored coefficient in a Pooling window, as illustrated in \autoref{fig:SI_Pooling}.

Consider a $2$-dimensional input tensor $x \in \mathbb{R}^{d_1 \times d_2}$ tensor encoded using a photonic circuit in a quantum state $\ket{x} = \sum_{i=1}^{d_1} \sum_{j=1}^{d_2} \frac{x_{i,j}}{||x||_2} \ket{e_{d_1,i}, e_{d_2,j}}$, with $\ket{e_{d_l, i_l}} = \ket{0 \dots 0 1 0 \dots 0}$ is a Fock state corresponding to no photon in $d_l$ modes except for the mode $i_l$ with a single particle. For $d_1 = d_2 = 4$, the Pooling layer applied on both registers outputs a state described by the following density operator:
\begin{equation}\label{eq:Pooling_Density_operator}
    \ket{x}\bra{x} \rightarrow 
        \begin{pmatrix} x^2_{11} + x^2_{12} + x^2_{13} + x^2_{14} & x_{12} x_{14} + x_{22} x_{24} & x_{21} x_{41} + x_{22} x_{42} & x_{22} x_{44} \\
        x_{12} x_{14} + x_{22} x_{24} & x^2_{13} + x^2_{14} + x^2_{23} + x^2_{24} & x_{24} x_{42} & x_{23} x_{43} + x_{24} x_{44} \\
        x_{21} x_{41} + x_{22} x_{42} & x_{24} x_{42} & x^2_{31} + x^2_{32} + x^2_{41} + x^2_{42} & x_{32} x_{34} + x_{42} x_{44} \\
        x_{22} x_{44} & x_{23} x_{43} + x_{24} x_{44} & x_{32} x_{34} + x_{42} x_{44} & x^2_{33} + x^2_{34} + x^2_{43} + x^2_{44}
        \end{pmatrix} \, \textrm{.}
\end{equation}
with $\ket{x}\bra{x} = \text{diag}(x_{11}^2, x_{12}^2, \dots, x_{44}^2)$. The value of each new coefficients is given by a polynomial combination of values from the corresponding Pooling window, i.e., a set of 4 neighbored coefficients in the input tensor. Notice that the probability of measuring the output state in state $\ket{e_i, e_j}$ corresponds to the sum of the probabilities of measuring the state corresponding to the coefficients in the input tensor. As a result, the produced tensor is very similar to an average pooling as explained in \cite{monbroussou2024subspace}, with a quadratic non-linearity on the produced density operator due to the measurement based nature of the layer.

\begin{figure}[h!]
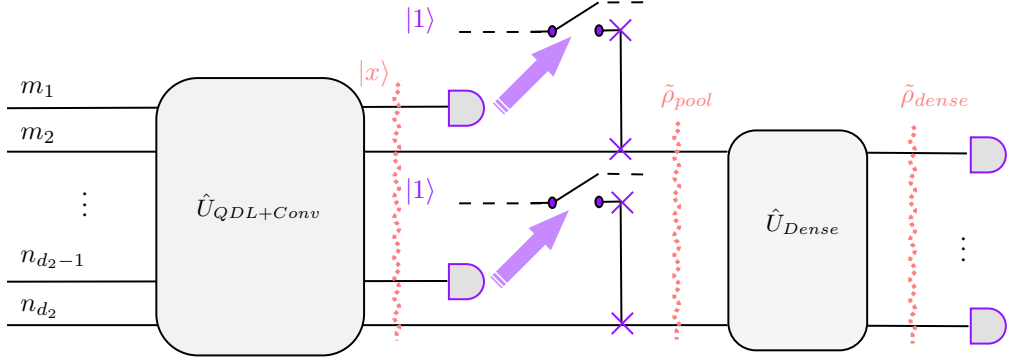

    \centering
    \include{Figures/figureRealSI}
    \caption{Scheme of state injection without postselection.}
    \label{fig:RealSI}
\end{figure}

To gain a clearer understanding of the optical resources required for the implementation of the \textbf{PQCNN} scheme without post-selection, as detailed in \autoref{subsec:ModularAdaptive}, we refer to \autoref{fig:RealSI} which is a zoom in \autoref{fig:exp_setup}a. The scenario would begin with ($k$) input photons across ($m=\sum_{i=1}^k d_i$) modes on a chip, where the QDL and Convolutional layers are put into operation. Then, for the pooling layer, we transition from one integrated interferometer to another, utilizing bulk optics, measuring ($m/2$) modes \footnote{In the applications presented in this article we were limited to even dimension registers and then measuring half of the modes and thus halving the dimension of the the dense compared to the input modes, but in general, one could imagine different pooling functions and/or odd register dimensions}, and injecting a maximum of ($k-1$) new photons into the unmeasured modes before moving on to the next integrated interferometer, where the dense layer is executed.\\

The total number of photons throughout all layers would be ($2k-1$), ideally, if multiple indistinguishable sources were used, the rate would be limited to the slowest emission. Otherwise, if one source was used followed by a DMX the emmision rate would be scaling as $(\eta_{total})^k$ where $\eta_{total}$ being the total efficiency of emission, switching and transmission of the demultiplexer, so it will scale exponentially with the number of photons $k$~\cite{pont2022high}.\\

As for the adaptivity time, it will depend on the time of measurements followed by electro-optical switching.  \\

It is also worth mentioning the depth of the optical circuit, even though it does not affect the running time of an optical circuit, we define the depth as the number of layers of Beam splitters that connect all nearest neighbors of all optical modes. For the first part of QDL+ Convolutional we get a depth of $O(K)$ since the depth will depend on the register with the highest dimension ($K$). Similarly for the dense layer it will be limited by the reduced number of mode and the number of parameters needed for the dense layer. As for the pooling layer it is sufficient to consider one layer of ($m/2$) switches so that one can inject photon in the next unmeasured mode if a photon was detected in the measured one. 

%% file: Figures/figureRealSI.tex
\tikzset{every picture/.style={line width=0.75pt}} 

\begin{tikzpicture}[x=0.75pt,y=0.75pt,yscale=-1,xscale=1]

\draw [line width=0.75]    (407.99,110.63) -- (528.64,111.52) ;
\draw [line width=0.75]    (407.99,198.18) -- (529.44,198.72) ;
\draw [color={rgb, 255:red, 255; green, 127; blue, 127 }  ,draw opacity=1 ][line width=1.5]  [dash pattern={on 1.5pt off 1.5pt}]  (501.26,97.77) .. controls (502.93,99.44) and (502.93,101.1) .. (501.26,102.77) .. controls (499.59,104.44) and (499.59,106.1) .. (501.26,107.77) .. controls (502.93,109.44) and (502.93,111.1) .. (501.26,112.77) .. controls (499.59,114.44) and (499.59,116.1) .. (501.26,117.77) .. controls (502.93,119.44) and (502.93,121.1) .. (501.26,122.77) .. controls (499.59,124.44) and (499.59,126.1) .. (501.26,127.77) .. controls (502.93,129.44) and (502.93,131.1) .. (501.26,132.77) .. controls (499.59,134.44) and (499.59,136.1) .. (501.26,137.77) .. controls (502.93,139.44) and (502.93,141.1) .. (501.26,142.77) .. controls (499.59,144.44) and (499.59,146.1) .. (501.26,147.77) .. controls (502.93,149.44) and (502.93,151.1) .. (501.26,152.77) .. controls (499.59,154.44) and (499.59,156.1) .. (501.26,157.77) .. controls (502.93,159.44) and (502.93,161.1) .. (501.26,162.77) .. controls (499.59,164.44) and (499.59,166.1) .. (501.26,167.77) .. controls (502.93,169.44) and (502.93,171.1) .. (501.26,172.77) .. controls (499.59,174.44) and (499.59,176.1) .. (501.26,177.77) .. controls (502.93,179.44) and (502.93,181.1) .. (501.26,182.77) .. controls (499.59,184.44) and (499.59,186.1) .. (501.26,187.77) .. controls (502.93,189.44) and (502.93,191.1) .. (501.26,192.77) .. controls (499.59,194.44) and (499.59,196.1) .. (501.26,197.77) .. controls (502.93,199.44) and (502.93,201.1) .. (501.26,202.77) -- (501.26,207.47) -- (501.26,207.47) ;
\draw [line width=0.75]    (44.52,88.69) -- (267.1,87.9) ;
\draw [line width=0.75]    (44.52,110.63) -- (407.77,110.63) ;
\draw  [color={rgb, 255:red, 144; green, 19; blue, 254 }  ,draw opacity=1 ][fill={rgb, 255:red, 155; green, 155; blue, 155 }  ,fill opacity=0.3 ][line width=0.75]  (267.38,79.82) -- (276.23,79.82) .. controls (281.12,79.82) and (285.08,83.79) .. (285.08,88.69) .. controls (285.08,93.58) and (281.12,97.55) .. (276.23,97.55) -- (267.38,97.55) -- cycle ;
\draw [line width=0.75]  [dash pattern={on 4.5pt off 4.5pt}]  (272.54,50.21) -- (313.9,50.21) ;
\draw [line width=0.75]    (44.52,176.25) -- (266.6,175.9) ;
\draw [line width=0.75]    (44.52,198.18) -- (407.77,198.18) ;
\draw  [color={rgb, 255:red, 144; green, 19; blue, 254 }  ,draw opacity=1 ][fill={rgb, 255:red, 155; green, 155; blue, 155 }  ,fill opacity=0.3 ][line width=0.75]  (267.38,167.38) -- (276.23,167.38) .. controls (281.12,167.38) and (285.08,171.34) .. (285.08,176.24) .. controls (285.08,181.14) and (281.12,185.1) .. (276.23,185.1) -- (267.38,185.1) -- cycle ;
\draw [color={rgb, 255:red, 255; green, 127; blue, 127 }  ,draw opacity=1 ][line width=1.5]  [dash pattern={on 1.5pt off 1.5pt}]  (382.26,95.77) .. controls (383.93,97.44) and (383.93,99.1) .. (382.26,100.77) .. controls (380.59,102.44) and (380.59,104.1) .. (382.26,105.77) .. controls (383.93,107.44) and (383.93,109.1) .. (382.26,110.77) .. controls (380.59,112.44) and (380.59,114.1) .. (382.26,115.77) .. controls (383.93,117.44) and (383.93,119.1) .. (382.26,120.77) .. controls (380.59,122.44) and (380.59,124.1) .. (382.26,125.77) .. controls (383.93,127.44) and (383.93,129.1) .. (382.26,130.77) .. controls (380.59,132.44) and (380.59,134.1) .. (382.26,135.77) .. controls (383.93,137.44) and (383.93,139.1) .. (382.26,140.77) .. controls (380.59,142.44) and (380.59,144.1) .. (382.26,145.77) .. controls (383.93,147.44) and (383.93,149.1) .. (382.26,150.77) .. controls (380.59,152.44) and (380.59,154.1) .. (382.26,155.77) .. controls (383.93,157.44) and (383.93,159.1) .. (382.26,160.77) .. controls (380.59,162.44) and (380.59,164.1) .. (382.26,165.77) .. controls (383.93,167.44) and (383.93,169.1) .. (382.26,170.77) .. controls (380.59,172.44) and (380.59,174.1) .. (382.26,175.77) .. controls (383.93,177.44) and (383.93,179.1) .. (382.26,180.77) .. controls (380.59,182.44) and (380.59,184.1) .. (382.26,185.77) .. controls (383.93,187.44) and (383.93,189.1) .. (382.26,190.77) .. controls (380.59,192.44) and (380.59,194.1) .. (382.26,195.77) .. controls (383.93,197.44) and (383.93,199.1) .. (382.26,200.77) -- (382.26,205.47) -- (382.26,205.47) ;
\draw [color={rgb, 255:red, 255; green, 127; blue, 127 }  ,draw opacity=1 ][line width=1.5]  [dash pattern={on 1.5pt off 1.5pt}]  (241.06,76.14) .. controls (242.73,77.81) and (242.73,79.47) .. (241.06,81.14) .. controls (239.39,82.81) and (239.39,84.47) .. (241.06,86.14) .. controls (242.73,87.81) and (242.73,89.47) .. (241.06,91.14) .. controls (239.39,92.81) and (239.39,94.47) .. (241.06,96.14) .. controls (242.73,97.81) and (242.73,99.47) .. (241.06,101.14) .. controls (239.39,102.81) and (239.39,104.47) .. (241.06,106.14) .. controls (242.73,107.81) and (242.73,109.47) .. (241.06,111.14) .. controls (239.39,112.81) and (239.39,114.47) .. (241.06,116.14) .. controls (242.73,117.81) and (242.73,119.47) .. (241.06,121.14) .. controls (239.39,122.81) and (239.39,124.47) .. (241.06,126.14) .. controls (242.73,127.81) and (242.73,129.47) .. (241.06,131.14) .. controls (239.39,132.81) and (239.39,134.47) .. (241.06,136.14) -- (241.06,136.7) -- (241.06,136.7) .. controls (242.73,138.37) and (242.73,140.03) .. (241.06,141.7) .. controls (239.39,143.37) and (239.39,145.03) .. (241.06,146.7) .. controls (242.73,148.37) and (242.73,150.03) .. (241.06,151.7) .. controls (239.39,153.37) and (239.39,155.03) .. (241.06,156.7) .. controls (242.73,158.37) and (242.73,160.03) .. (241.06,161.7) .. controls (239.39,163.37) and (239.39,165.03) .. (241.06,166.7) .. controls (242.73,168.37) and (242.73,170.03) .. (241.06,171.7) .. controls (239.39,173.37) and (239.39,175.03) .. (241.06,176.7) .. controls (242.73,178.37) and (242.73,180.03) .. (241.06,181.7) .. controls (239.39,183.37) and (239.39,185.03) .. (241.06,186.7) .. controls (242.73,188.37) and (242.73,190.03) .. (241.06,191.7) .. controls (239.39,193.37) and (239.39,195.03) .. (241.06,196.7) .. controls (242.73,198.37) and (242.73,200.03) .. (241.06,201.7) -- (241.06,205.97) -- (241.06,205.97) ;
\draw  [fill={rgb, 255:red, 244; green, 241; blue, 241 }  ,fill opacity=1 ] (119.83,94.42) .. controls (119.83,82.85) and (129.21,73.47) .. (140.79,73.47) -- (203.65,73.47) .. controls (215.22,73.47) and (224.6,82.85) .. (224.6,94.42) -- (224.6,192.51) .. controls (224.6,204.09) and (215.22,213.47) .. (203.65,213.47) -- (140.79,213.47) .. controls (129.21,213.47) and (119.83,204.09) .. (119.83,192.51) -- cycle ;
\draw  [fill={rgb, 255:red, 247; green, 244; blue, 244 }  ,fill opacity=1 ] (408.33,113.9) .. controls (408.33,106.17) and (414.6,99.9) .. (422.33,99.9) -- (464.33,99.9) .. controls (472.07,99.9) and (478.33,106.17) .. (478.33,113.9) -- (478.33,196.9) .. controls (478.33,204.63) and (472.07,210.9) .. (464.33,210.9) -- (422.33,210.9) .. controls (414.6,210.9) and (408.33,204.63) .. (408.33,196.9) -- cycle ;
\draw  [color={rgb, 255:red, 144; green, 19; blue, 254 }  ,draw opacity=1 ][fill={rgb, 255:red, 155; green, 155; blue, 155 }  ,fill opacity=0.3 ][line width=0.75]  (530.58,103.38) -- (539.43,103.38) .. controls (544.32,103.38) and (548.28,107.34) .. (548.28,112.24) .. controls (548.28,117.14) and (544.32,121.1) .. (539.43,121.1) -- (530.58,121.1) -- cycle ;
\draw  [color={rgb, 255:red, 144; green, 19; blue, 254 }  ,draw opacity=1 ][fill={rgb, 255:red, 155; green, 155; blue, 155 }  ,fill opacity=0.3 ][line width=0.75]  (530.58,189.78) -- (539.43,189.78) .. controls (544.32,189.78) and (548.28,193.74) .. (548.28,198.64) .. controls (548.28,203.54) and (544.32,207.5) .. (539.43,207.5) -- (530.58,207.5) -- cycle ;
\draw  [fill={rgb, 255:red, 144; green, 19; blue, 254 }  ,fill opacity=1 ] (308.65,50.1) -- (317.71,49.94) (344.91,49.46) -- (353.98,49.3) (321.32,48.99) -- (342.85,35.3) (341.28,49.53) .. controls (341.26,48.06) and (342.05,46.85) .. (343.05,46.83) .. controls (344.05,46.81) and (344.89,47.99) .. (344.91,49.46) .. controls (344.94,50.93) and (344.15,52.14) .. (343.14,52.16) .. controls (342.14,52.17) and (341.31,51) .. (341.28,49.53) -- cycle (317.71,49.94) .. controls (317.69,48.47) and (318.48,47.26) .. (319.48,47.24) .. controls (320.48,47.23) and (321.31,48.4) .. (321.34,49.87) .. controls (321.37,51.34) and (320.57,52.55) .. (319.57,52.57) .. controls (318.57,52.58) and (317.74,51.41) .. (317.71,49.94) -- cycle ;
\draw [line width=0.75]  [dash pattern={on 4.5pt off 4.5pt}]  (347.35,35.3) -- (370.9,35.2) ;
\draw    (353.98,49.3) -- (354.4,110.7) ;
\draw [line width=0.75]  [dash pattern={on 4.5pt off 4.5pt}]  (272.54,136.71) -- (313.9,136.71) ;
\draw  [fill={rgb, 255:red, 144; green, 19; blue, 254 }  ,fill opacity=1 ] (308.65,136.6) -- (317.71,136.44) (344.91,135.96) -- (353.98,135.8) (321.32,135.49) -- (342.85,121.8) (341.28,136.03) .. controls (341.26,134.56) and (342.05,133.35) .. (343.05,133.33) .. controls (344.05,133.31) and (344.89,134.49) .. (344.91,135.96) .. controls (344.94,137.43) and (344.15,138.64) .. (343.14,138.66) .. controls (342.14,138.67) and (341.31,137.5) .. (341.28,136.03) -- cycle (317.71,136.44) .. controls (317.69,134.97) and (318.48,133.76) .. (319.48,133.74) .. controls (320.48,133.73) and (321.31,134.9) .. (321.34,136.37) .. controls (321.37,137.84) and (320.57,139.05) .. (319.57,139.07) .. controls (318.57,139.08) and (317.74,137.91) .. (317.71,136.44) -- cycle ;
\draw [line width=0.75]  [dash pattern={on 4.5pt off 4.5pt}]  (347.35,121.8) -- (369.4,122.2) ;
\draw    (353.98,135.8) -- (354.4,197.2) ;
\draw  [color={rgb, 255:red, 0; green, 0; blue, 0 }  ,draw opacity=0 ][fill={rgb, 255:red, 144; green, 19; blue, 254 }  ,fill opacity=0.5 ] (291.1,171.18) -- (310.37,151.97) -- (307.56,149.15) -- (327.92,140.11) -- (318.81,160.43) -- (316,157.61) -- (296.72,176.83) -- cycle ;\draw  [color={rgb, 255:red, 0; green, 0; blue, 0 }  ,draw opacity=0 ][fill={rgb, 255:red, 144; green, 19; blue, 254 }  ,fill opacity=0.5 ] (288.28,173.99) -- (288.84,173.43) -- (294.47,179.08) -- (293.9,179.64) -- cycle ;\draw  [color={rgb, 255:red, 0; green, 0; blue, 0 }  ,draw opacity=0 ][fill={rgb, 255:red, 144; green, 19; blue, 254 }  ,fill opacity=0.5 ] (289.4,172.87) -- (290.53,171.74) -- (296.16,177.39) -- (295.03,178.51) -- cycle ;
\draw  [color={rgb, 255:red, 0; green, 0; blue, 0 }  ,draw opacity=0 ][fill={rgb, 255:red, 144; green, 19; blue, 254 }  ,fill opacity=0.5 ] (292.43,83.51) -- (311.7,64.3) -- (308.89,61.48) -- (329.25,52.44) -- (320.15,72.77) -- (317.33,69.95) -- (298.06,89.16) -- cycle ;\draw  [color={rgb, 255:red, 0; green, 0; blue, 0 }  ,draw opacity=0 ][fill={rgb, 255:red, 144; green, 19; blue, 254 }  ,fill opacity=0.5 ] (289.61,86.33) -- (290.17,85.77) -- (295.8,91.41) -- (295.24,91.97) -- cycle ;\draw  [color={rgb, 255:red, 0; green, 0; blue, 0 }  ,draw opacity=0 ][fill={rgb, 255:red, 144; green, 19; blue, 254 }  ,fill opacity=0.5 ] (290.74,85.2) -- (291.87,84.08) -- (297.49,89.72) -- (296.36,90.85) -- cycle ;
\draw  [color={rgb, 255:red, 144; green, 19; blue, 254 }  ,draw opacity=1 ] (349.51,191.91) -- (360.28,202.68)(360.28,191.91) -- (349.51,202.68) ;
\draw  [color={rgb, 255:red, 144; green, 19; blue, 254 }  ,draw opacity=1 ] (349.51,130.91) -- (360.28,141.68)(360.28,130.91) -- (349.51,141.68) ;
\draw  [color={rgb, 255:red, 144; green, 19; blue, 254 }  ,draw opacity=1 ] (348.51,43.91) -- (359.28,54.68)(359.28,43.91) -- (348.51,54.68) ;
\draw  [color={rgb, 255:red, 144; green, 19; blue, 254 }  ,draw opacity=1 ] (348.51,103.91) -- (359.28,114.68)(359.28,103.91) -- (348.51,114.68) ;

\draw (493.73,75) node [anchor=north west][inner sep=0.75pt]  [font=\normalsize,color={rgb, 255:red, 255; green, 127; blue, 127 }  ,opacity=1 ] [align=left] {$\displaystyle \tilde{\rho }_{dense}$};
\draw (243.83,35.98) node [anchor=north west][inner sep=0.75pt]  [font=\normalsize] [align=left] {$\displaystyle \textcolor[rgb]{0.56,0.07,1}{\ket{1}}$};
\draw (49.87,74) node [anchor=north west][inner sep=0.75pt]  [font=\normalsize] [align=left] {$\displaystyle m_{1}$};
\draw (49.87,97) node [anchor=north west][inner sep=0.75pt]  [font=\normalsize] [align=left] {$\displaystyle m_{2}$};
\draw (49.87,159) node [anchor=north west][inner sep=0.75pt]  [font=\normalsize] [align=left] {$\displaystyle n_{d_{2} -1}$};
\draw (49.87,184) node [anchor=north west][inner sep=0.75pt]  [font=\normalsize] [align=left] {$\displaystyle n_{d_{2}}$};
\draw (80.6,121.8) node [anchor=north west][inner sep=0.75pt]  [font=\normalsize] [align=left] {$\displaystyle \vdots $};
\draw (372.73,75) node [anchor=north west][inner sep=0.75pt]  [font=\normalsize,color={rgb, 255:red, 255; green, 127; blue, 127 }  ,opacity=1 ] [align=left] {$\displaystyle \tilde{\rho }_{pool}$};
\draw (220,63) node [anchor=north west][inner sep=0.75pt]  [font=\normalsize,color={rgb, 255:red, 255; green, 127; blue, 127 }  ,opacity=1 ] [align=left] {$\displaystyle \ket{x}$};
\draw (523.8,144.2) node [anchor=north west][inner sep=0.75pt]  [font=\normalsize] [align=left] {$\displaystyle \vdots $};
\draw (133,130) node [anchor=north west][inner sep=0.75pt]   [align=left] {\begin{minipage}[lt]{55.73pt}\setlength\topsep{0pt}
\begin{center}
 $\displaystyle \hat{U}_{QDL+Conv}$
\end{center}

\end{minipage}};
\draw (420.6,138) node [anchor=north west][inner sep=0.75pt]   [align=left] {\begin{minipage}[lt]{35.78pt}\setlength\topsep{0pt}
\begin{center}
 $\displaystyle \hat{U}_{Dense}$
\end{center}

\end{minipage}};
\draw (243.83,122.48) node [anchor=north west][inner sep=0.75pt]  [font=\normalsize] [align=left] {$\displaystyle \textcolor[rgb]{0.56,0.07,1}{\ket{1}}$};

\end{tikzpicture}

%% file: Supplementary_Materials/Appendix_Simulations.tex
\section{Simulation Results}\label{chap:Simulation_Results}

\begin{figure*}
    \centering
    \includegraphics[width=\textwidth]{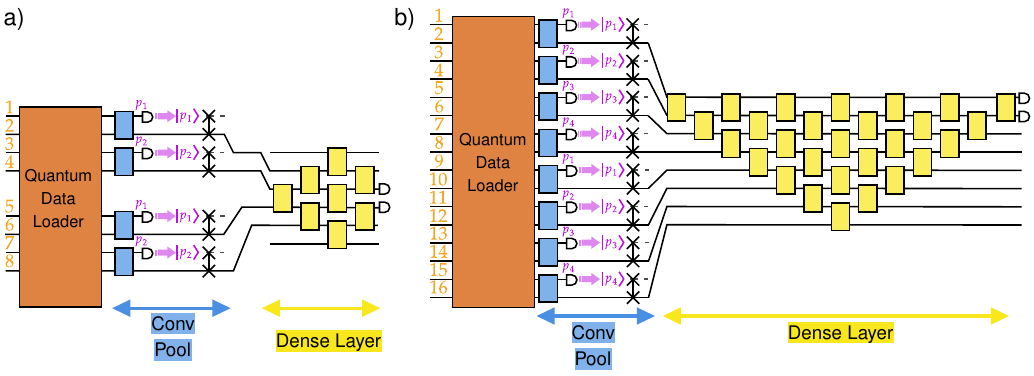}
    \caption{\textbf{Simulated architecture. a)} the $8$ mode architecture corresponding to the experimental apparatus is used to classify both BAS datasets. \textbf{b)} the $16$ mode architecture is used to classify the MNIST dataset.}
    \label{fig:simulated architecture}
\end{figure*}

In this Section, we give more details about the simulations performed to obtain \autoref{table:Simulation_Classification}. For each dataset, we consider $5$ random initial set of parameters. Each method is trained through $30$ epochs using ADAM optimizer, a learning rate $\gamma=10^{-3}$, and a weight decay $\lambda=10^{-4}$. The first two datasets are trained and simulated considering the architecture described in \autoref{sec:Experimental_Apparatus}.

\begin{figure*}
    \centering
    \includegraphics[width=\textwidth]{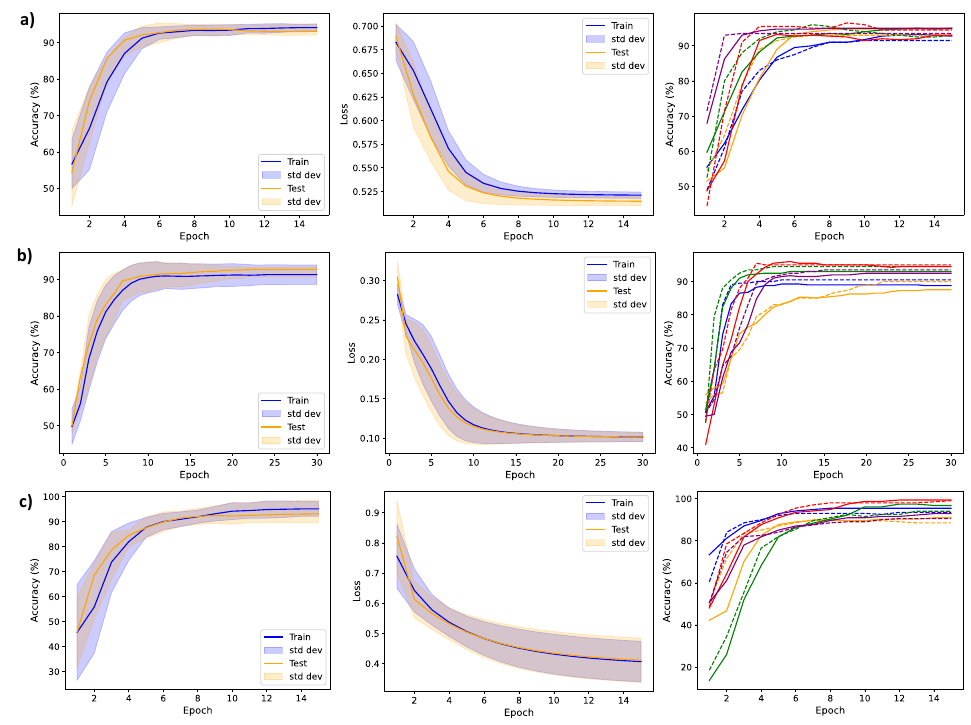}
    \caption{\textbf{Simulation Results. a)} BAS dataset accuracy, loss, and all accuracies. \textbf{b)} Custom BAS dataset accuracy, loss, and all accuracies. \textbf{c)} MNIST dataset accuracy, loss, and all accuracies.}
    \label{fig:subfigures}
\end{figure*}

The different architectures used in the simulation are presented in \autoref{fig:simulated architecture}.

%% file: Supplementary_Materials/Details_setup.tex
\section{Details about the integrated circuit and the experimental setup}

\begin{figure}[h!]
    \centering
    \includegraphics[width=0.9\linewidth]{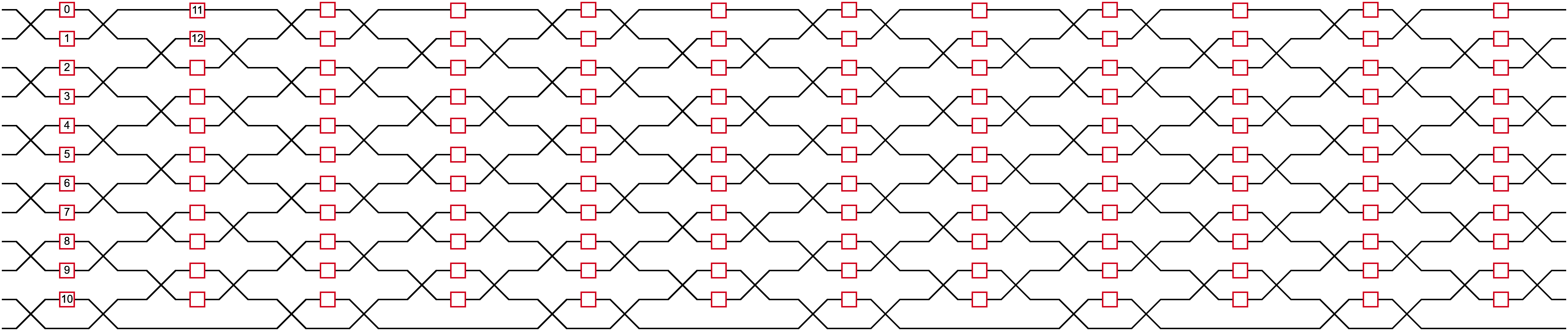}
    \caption{\textbf{Circuital scheme of the waveguides and thermo-optic phase shifters for the 12-mode fully reconfigurable device.} The Black lines identify the waveguides, while the red rectangles highlight the position of the thermo-optic phase shifters.}
    \label{fig:12modi}
\end{figure}

The twelve-mode integrated circuit in our experiment features a waveguide optical layout based on the "universal" design reported in \cite{Clements:16}, and is made programmable through thermo-optical phase shifters. The detailed layout of the circuit is illustrated in \autoref{fig:12modi}.

The core of the circuit is a rectangular mesh of Mach-Zehnder interferometers (MZIs), each composed of two cascaded directional couplers, which are brought into proximity in the central region to enable optical power exchange via evanescent-field interaction. Each MZI has two thermo-optic phase shifters, as shown in \autoref{fig:12modi}. The lateral pitch between the interferometer arms is {80} $\mu$m. The Fan-in and Fan-out sections are integrated at both ends of the circuit to match the 127 $\mu$m pitch of standard fiber arrays. The entire circuit has been implemented on a glass device measuring {15 x 90 $mm^2$}.
The photonic device was fabricated on a commercial alumino-borosilicate glass substrate (Eagle XG) with a thickness of 1.1 mm, using femtosecond laser writing technology. Specifically, we employed a Yb-based laser system (Light conversion Pharos), which delivers pulse trains at $\lambda=1030$ nm, with a tunable repetition rate ranging from 1kHz to 1MHz, a pulse duration of 170 fs, and an average power of up to 10 W.
The waveguides were inscribed 25 $\mu$m below the top surface of the glass substrate. During the inscription process, the laser operated at a 1 MHz repetition rate with 330nJ pulse energy. Focusing was achieved using a 20x objective (0.5 NA), while the substrate was translated at a constant speed of 25.0 mm/s. Following irradiation, the substrate underwent a thermal annealing treatment, enhancing the optical confinement and reducing propagation losses. The total optical insertion loss of the circuit was measured to be 3 dB. Thermo-optical phase shifters consist of resistive microheaters deposited on the top surface of the chip device, precisely aligned above the waveguide regions where a tunable phase shift is required. In particular,  driving a controlled current into the microheater, a controlled temperature increase is achieved in a localized fashion, thus inducing a thermo-optic effect, which is a phase delay in the light propagating through the waveguide.\\

\textbf{Single-photon source, photon synchronization, and detection.}\\

\begin{figure}[H]
    \centering
    \includegraphics[width=0.9\linewidth]{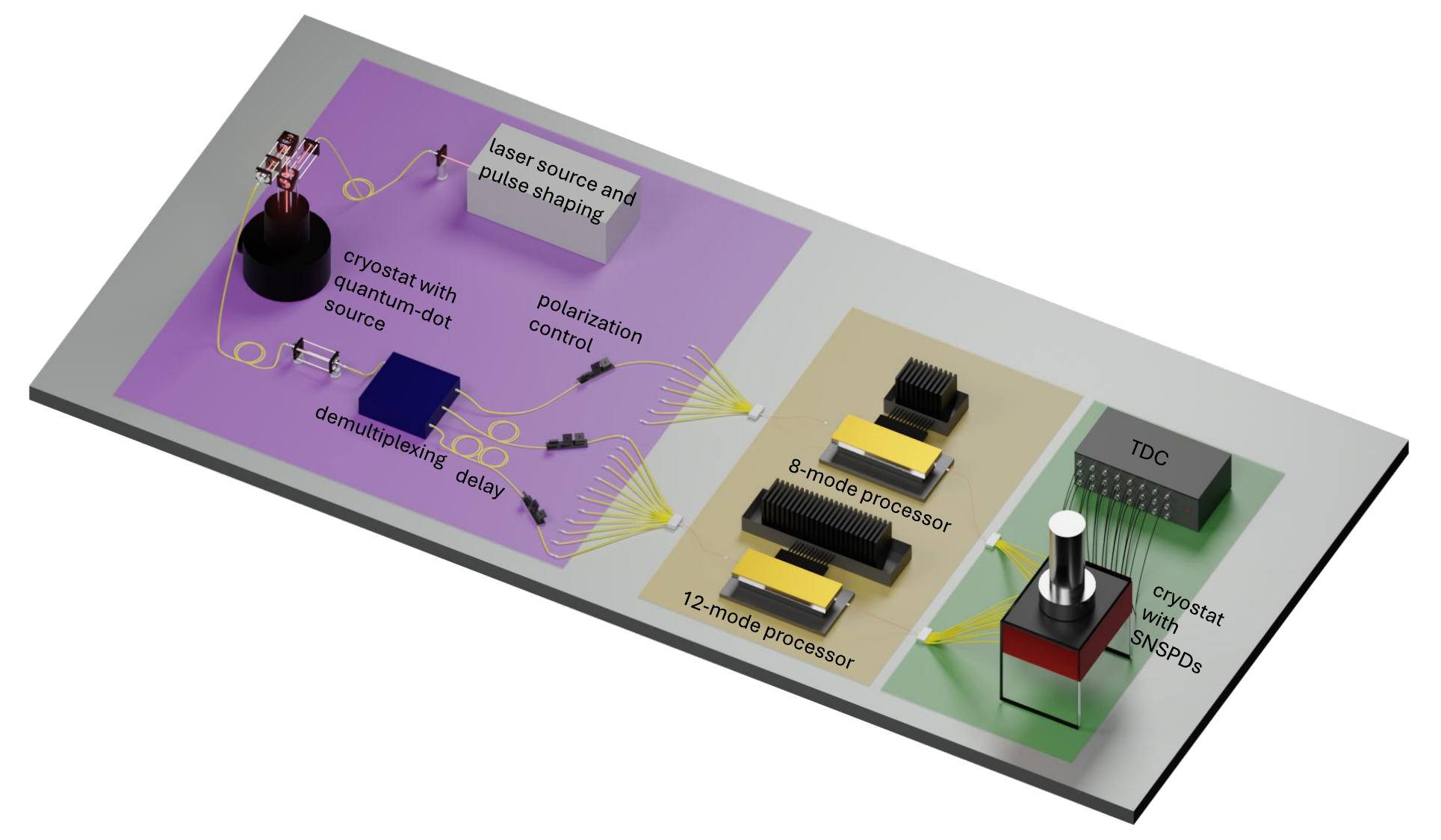}
    \caption{\textbf{Scheme of the complete experimental setup.} The quantum dot source emits a train of single photons, which are distributed in different spatial modes by a time-to-spatial demultiplexing system, subsequently temporally synchronized via fiber delays and are prepared in the same polarization state. The photons are coupled in two photonic fully-reconfigurable integrated interferometers, having respectively 8 and 12 spatial modes with universal programming capabilities. The output modes of the two processor are coupled to a system of SNSPDs, whose output signals are analyzed in a time-to-digital converter (TDC).}
    \label{fig:full_setup}
\end{figure}

The experimental platform shown in \autoref{fig:full_setup} exploits photons generated on-demand by a quantum dot emitter. The key characteristic of this kind of source is the high brightness, which enables the implementation of multiphoton experiments with near-deterministic performance. In particular, the quantum dot emitter used in this experiment is a commercial solution provided by Quandela, operating at a cryogenic temperature of 4 K. The emitter consists of a single self-assembled InGaAs quantum dot embedded within an electrically controlled microcavity. The system's source operates in the resonant regime (RF). \cite{somaschi2016near,ollivier2020reproducibility,nowak2014deterministic,gazzano2013bright}, i.e. the pulsed pump laser excites the source matching its excitonic transition energy, corresponding to an optical wavelength of approximately $\sim 928$ nm, and residual
laser filtering is obtained via a typical cross-polarization scheme. Moreover, to increase the generation rate, the repetition rate of the laser before the quantum dot emitter is 160 MHz thus allowing a generation count rate of 10 Mhz when measured through Superconductive Nanowire Single-Photon Detectors (SNSPDs) of the Single-Quantum Company. The typical single-photon purity - measured via a typical Hanbury-Brown-Twiss setup - is typically around $\mathcal{P} = 1 - g^{(2)}(0) \sim 0.97$. Conversely, the pairwise indistinguishability between subsequently emitted single-photons - quantified via the Hong-Ou-Mandel visibility measured within a time-unbalanced Mach-Zehnder interferometer has typical values of $V$ between $0.83-0.90$.

The quantum dot source is then interfaced with a time-to-spatial demultiplexing (DMX) module, based on an acousto-optic modulator \cite{rodari2024semi,rodari2024experimental, hansen2023single,pont2022quantifying,pont2022high}. This module transforms the initial time-separated sequence of single photons into a set of photons distributed across multiple spatial modes to effectively create a multiphoton resource that can be injected into the input ports of the integrated devices. The visibility of the Hong-Ou-Mandel (HOM) interference for the three possible photon pairs among the photon pairs entering the 12-mode integrated interferometer, as measured at the output of the circuit, has a value close to the HOM visibility measured at the output of the single-photon source. The third photon, injected into the 8-mode device, evolves without interfering with other photons. The outgoing photons are then detected by a Superconducting Nanowire Single-Photon Detectors (SNSPDs) system. A Time-to-Digital Converter (TDC) registers the manifold coincidences between photons.